\documentclass[aps,prb,reprint,groupedaddress,amsmath,amsfonts,amssymb,floatfix]{revtex4-1}

\usepackage{xcolor}
\definecolor{darkgreen}{rgb}{0,0.6,0}
\definecolor{cyan}{rgb}{0,0.7,0.8}
\usepackage[colorlinks=true,citecolor=darkgreen,urlcolor=cyan]{hyperref}
\usepackage{graphicx}
\usepackage{multirow}
\usepackage{diagbox}

\newcommand{\mrm}[1]{\mathrm{#1}}

\newcommand{\mbb}[1]{\mathbb{#1}}
\newcommand{\mc}[1]{\mathcal{#1}}

\newcommand{\eref}[1]{(\ref{#1})}
\newcommand{\Eref}[1]{Eq.~(\ref{#1})}

\newcommand{\fref}[1]{Fig.~\ref{#1}}

\newcommand{\tref}[1]{Table~\ref{#1}}
\newcommand{\sref}[1]{Sec.~\ref{#1}}

\newcommand{\srefs}[2]{Secs.~\ref{#1} and~\ref{#2}}

\newcommand{\pEref}[1]{\protect{Eq.~(\ref{#1})}}
\newcommand{\pfref}[1]{\protect{Fig.~\ref{#1}}}

\newcommand{\rmd}{\mathrm{d}}

\newcommand{\ra}{\rangle}
\newcommand{\la}{\langle}

\newcommand{\rcite}[1]{Ref.~\onlinecite{#1}}

\newcommand{\bonderson}{Refs.~\onlinecite{bonderson2007} and \onlinecite{bonderson2008}}
\newcommand{\MATLAB}{\textsc{Matlab}}

\begin{document}

\title{Finite Density Matrix Renormalisation Group Algorithm for Anyonic Systems}

\author{Robert N. C. Pfeifer}
\email[]{robert.pfeifer@mq.edu.au}
\affiliation{Dept. of Physics \& Astronomy, Macquarie University, Sydney, NSW 2109, Australia}
\author{Sukhwinder Singh}
\affiliation{Center for Engineered Quantum Systems; Dept. of Physics \& Astronomy, Macquarie University, 2109 NSW, Australia}

\date{\today}

\begin{abstract}
The numerical study of anyonic systems is known to be highly challenging due to their non-bosonic, non-fermionic particle exchange statistics, and with the exception of certain models for which analytical solutions exist, very little is known about their collective behaviour as a result. Meanwhile, the density matrix renormalisation group (DMRG) algorithm is an exceptionally powerful numerical technique for calculating the ground state of a low-dimensional lattice Hamiltonian, and %
has been applied to the study of bosonic, fermionic, and group-symmetric systems. The recent development of a tensor network formulation for anyonic systems opened up the possibility of studying these systems using algorithms such as DMRG, though this has proved challenging both in terms of programming complexity and computational cost. This paper presents the implementation of DMRG for finite anyonic systems, including a detailed scheme for the implementation of anyonic tensors with optimal scaling of computational cost. The anyonic DMRG algorithm is demonstrated by calculating the ground state energy of the Golden Chain, which has become the benchmark system for the numerical study of anyons, and is shown to produce results comparable to those of the anyonic TEBD algorithm and superior to the variationally optimised anyonic MERA, at far lesser computational cost.
\end{abstract}


\maketitle

\section{Introduction}

The Density Matrix Renormalisation Group (DMRG) algorithm\cite{white1992,white1993,schollwock2005} is an algorithm for computing the ground state of a low-dimensional lattice Hamiltonian, and is arguably one of the most successful and widespread algorithms in condensed matter physics.\cite{schollwock2005} With the advent of tensor network representations of quantum states, it is now recognised that DMRG %
may be understood as a variational tensor network algorithm %
to compute a Matrix Product State (MPS) representation of the ground state of a quantum system.\cite{verstraete2004,schollwock2011}

Anyons, meanwhile, are quasiparticles capable of existing only in one- or two-dimensional systems,\cite{kitaev2006} and are extremely challenging to simulate numerically due to their non-bosonic, non-fermionic exchange statistics. To date, simulation of anyonic systems has largely consisted of exact diagonalisation performed on chains or ladders of around 30-40 lattice sites,\cite{feiguin2007,trebst2008,poilblanc2011,poilblanc2012,pfeifer2012,pfeifer2012a} and exact calculations where mappings to analytically solvable models exist.\cite{feiguin2007,trebst2008a} 
Where available, such mappings have also been exploited to permit some systems with quantum group statistics to be studied using DMRG.\cite{sierra1997,tatsuaki2000,wada2001}
The study of anyonic systems has so far been largely theoretical, though the recently reported detection of Ising anyons at the ends of iron nanowires\cite{nadj-perge2014} may be about to change all that, lending pressing urgency to the development of efficient numerical algorithms for the simulation of anyonic systems. It would therefore be highly advantageous to convert existing, highly effective algorithms such as DMRG for use with arbitrary anyonic systems.

With this goal in mind, recent work on symmetries in tensor network states\cite{singh2010} led to the development of a tensor network formalism for anyonic systems.\cite{pfeifer2010,konig2010} This formalism permits existing tensor network algorithms to be adapted to the study of anyonic systems, including the scale-invariant Multi-Scale Entanglement Renormalisation Ansatz (MERA)\cite{vidal2007,vidal2008a,pfeifer2009,pfeifer2010} and the Time-Evolving Block Decimation (TEBD) algorithm.\cite{vidal2003,vidal2004,singh2014}
In the present paper, we show---in quite considerable detail---how the DMRG algorithm may also be implemented for anyonic systems. Our aims in doing so are twofold: First, how could we resist bringing together the most challenging particles to simulate, and the most powerful numerical technique for low-dimensional systems, especially when those particles can only exist in one or two dimensions? Second, the DMRG algorithm is the perfect context in which to provide a more detailed exposition of the anyonic tensor formalism first introduced in \rcite{pfeifer2010} and its application. Its tensor networks are more complex than those of TEBD but less so than those of the MERA, and so it is possible to present the algorithm at a level of detail which is simply not practical for the MERA, while nevertheless involving a level of sophistication which is not required for the simpler tensor networks encountered in TEBD.

This paper is organised as follows: 

In \sref{sec:tensorconstruction} we discuss the construction of anyonic tensors, the choice of bases on tensor indices, and how these relate to fusion tree diagrams. In \sref{sec:manipulation} we see how these tensors may be efficiently manipulated, and in \sref{sec:networks} we construct anyonic tensor networks, examine the pairwise contraction of anyonic tensors, and compare it with the whole-network approach employed in \rcite{singh2014}. We also compare and contrast the anyonic tensor network formalism with that used for conventional tensor networks. In \sref{sec:aMPS} we use anyonic tensors to define an MPS Ansatz for anyonic systems on the disc, and in %
\sref{sec:finiteADMRG} %
we show how the anyonic version of the finite DMRG algorithm uses this Ansatz to compute the ground state of a Hamiltonian %
on a finite chain on the disc. Section~\ref{sec:torus} briefly addresses the extension of this Ansatz to the study of chains on the torus as an example of anyonic DMRG on surfaces of higher genus. 

The infinite DMRG algorithm is not addressed in this paper: Its adaptation to anyonic systems is less straightforward than that of the finite DMRG algorithm, and caution is required to avoid inadvertently imposing non-physical constraints on the states which may be explored. The nature of these constraints and techniques for their avoidance are discussed in \rcite{pfeifer2015}, with their application to anyonic systems being addressed explicitly in Sec.~III~F of that paper.

\section{Anyonic tensors\label{sec:anyonictensors}}

This Section details the construction of anyonic tensors and tensor networks. A normalisation scheme is introduced which is compatible with the diagrammatic isotopy convention described in Refs.~\onlinecite{bonderson2007} and \onlinecite{bonderson2008}. Familiarity with the diagrammatic representation of anyonic states employed in these works is assumed. The focus of this Section is pragmatic, with emphasis on concerns such as choices of normalisation convention, and is intended to complement the more rigorous but also more abstract treatment presented in \rcite{pfeifer2010}. 

\subsection{Construction\label{sec:tensorconstruction}}

\subsubsection{Anyonic state vectors}

\begin{figure}
\begin{center}
\includegraphics[width=\columnwidth]{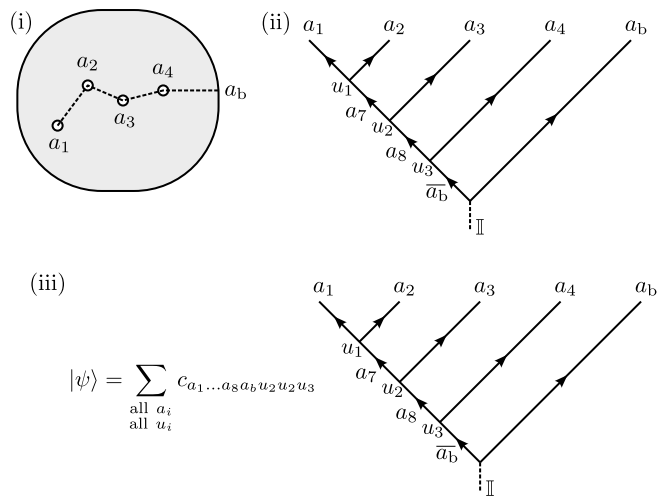}
\caption{Choosing a basis for a system of four anyons on the disc (plus a possible boundary charge): (i)~Four anyons on the disc. A linear ordering, represented by the dashed line, has been imposed. (ii)~A fusion tree is selected whose projection onto the disc coincides with the imposed linear ordering. The labels on this fusion tree represent a selection of combined charges. For example, $a_7$ is the total charge of sites~1 and~2 together, and $a_8$ is the total charge of sites~1, 2, and~3 together. (iii)~A state is constructed as a weighted sum over valid labelings of the fusion tree.\label{fig:intro}}
\end{center}
\end{figure}

As in \rcite{pfeifer2010} we begin by considering a surface of genus~0 on which there exists a lattice $\mc{L}$ of $n$ sites, populated by anyons [\fref{fig:intro}(i)]. 
We then choose a basis by introducing a linear ordering of these $n$ sites, as discussed in \rcite{pfeifer2014} and illustrated by the dashed line in \fref{fig:intro}(i), and choosing a unidirectional\footnote{By unidirectional, we mean a fusion tree made up entirely of fusion vertices, or entirely of splitting vertices.} fusion tree [\fref{fig:intro}(ii)], whose projection onto the manifold corresponds to the linear ordering adopted in \fref{fig:intro}(i). Extension of the fusion tree to the boundary of the disc permits inclusion of a boundary charge $a_b$. This charge may be represented in two ways: As a total charge at the bottom of the fusion tree, or as a leaf carrying charge $\overline{a_b}$. We favour the latter approach, as a fusion tree may then be understood as a recipe for the construction of a state, the lines of the tree corresponding to world lines, with cross-sections through the tree being edge-on views of the manifold at a given instant in time.%
\footnote{Although the boundary is an extended object and the charge $\overline{a_b}$ is point-like, the representation of the boundary charge by a leaf of the fusion tree is nevertheless fully general and rigorous: A lattice of \protect{$n$} sites on the disc with non-trivial boundary charge may be mapped to a lattice of \protect{$n$} sites on the sphere, and then to \protect{$n+1$} sites on the disc with trivial boundary charge, with the extra lattice site carrying the dual of the original boundary charge. This mapping is discussed in \protect{\rcite{pfeifer2012a}}.} Given the equivalence between a boundary charge and an additional lattice site, we will not need to explicitly consider the possibility of a boundary charge beyond this point.

A state is defined as a weighted sum over labellings of the fusion tree which are consistent with the anyonic fusion rules, and a total charge of $\mbb{I}$ [\fref{fig:intro}(iii)]. The number of indices involved may be reduced considerably by introducing an index $\mu_\mbb{I}$ which enumerates all labellings of the fusion tree consistent with these conditions. Any labelling of a fusion tree may then be identified with a pair $(a,\mu_a)$ where $a$ is the total charge of the fusion tree, and the admissible values of index $\mu_a$ enumerate the labellings of the tree consistent with total charge $a$. (For example, if there are seven valid labellings with a given value of total charge $a$, then $\mu_a$ takes an integer value in the range $[1,7]$.) When describing a state we will only need pairs for which $a=\mbb{I}$, but other values of $a$ will be necessary when describing operators and three-index tensors.

Using this more compact notation the coefficients describing a state may now be written
\begin{equation}
c^{(a,\mu_a)},
\end{equation}
and if we introduce a linear ordering of the pairs $(a,\mu_a)$ which is enumerated by greek index $\alpha$, then we obtain the state vector $c^\alpha$ where $\alpha$ enumerates the basis elements corresponding to \emph{all} valid labellings of the chosen fusion tree.

\begin{figure}
\begin{center}
\includegraphics[width=\columnwidth]{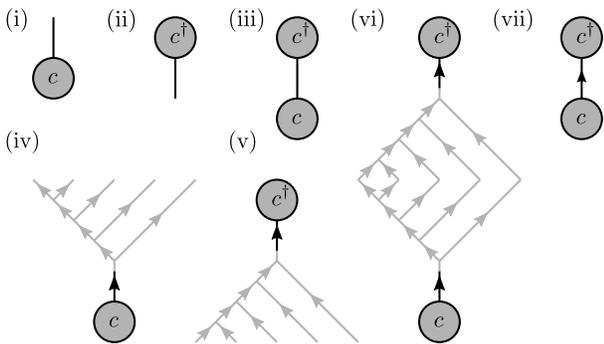}
\caption{(i)~Conventional tensor network notation for a vector $c^\alpha$ (i.e.~a single-index tensor), or associated state $|\psi\ra=\sum_\alpha c^\alpha |i_\alpha\ra$. The circle represents the tensor $c$, and the single emerging line represents the index. The conjugate state $\la\psi|$, and associated tensor $c^\dagger_\alpha$, is represented by diagram~(ii). (iii)~Diagram in conventional tensor network notation corresponding to the inner product $c^\alpha c^\dagger_\alpha$ (repeated index summed). The line connecting $c$ and $c^\dagger$ represents a summed index appearing on both tensors. (iv)~Extension of graphical notation for an anyonic vector $c^\alpha$, incorporating the fusion tree whose labellings are enumerated by $\alpha$. The fusion tree is drawn in grey to indicate a normalisation convention where no numerical factors are associated with vertices or loops.
(v)~The graphical representation of the conjugate state is obtained by vertical reflection and reversal of arrows. (vi)~The anyonic inner product, which reduces in this normalisation convention to diagram~(vii), again corresponding to $c^\alpha c^\dagger_\alpha$.\label{fig:vector}}
\end{center}
\end{figure}

For tensor network algorithms, it is conventional to represent a single-index tensor such as $c^\alpha$ by a circle (denoting the tensor) with a single line emerging (denoting the index), as shown in \fref{fig:vector}(i).
Ignoring, for the moment, that $\alpha$ corresponds to labellings of a fusion tree, and writing the basis of our Hilbert space as $|i_1\ra,\ldots,|i_\alpha\ra$, we may identify a vector $c^\alpha$ with a state
\begin{equation}
|\psi\ra=\sum_\alpha c^\alpha |i_\alpha\ra.
\end{equation}
Its conjugate,
\begin{equation}
\la\psi|=\sum_\alpha c^\dagger_\alpha \la i^\alpha|,\label{eq:hermconj}
\end{equation}
is obtained by taking the Hermitian conjugate of $c^\alpha$ to obtain $c^\dagger_\alpha$, i.e.
\begin{equation}
c^\dagger_\alpha = (c^\alpha)^*~\forall~\alpha,
\end{equation} 
vertically reflecting \fref{fig:vector}(i), and reversing all arrows (which is equivalent to conjugation of charges), to obtain \fref{fig:vector}(ii). Note that we write indices upper or lower to match the upgoing or downgoing orientations of the associated legs in the graphical notation.
It is convenient to have the vector representation of a physical state satisfy the normalisation condition 
\begin{equation}
\la\psi|\psi\ra=c^\alpha c^\dagger_\alpha=1\label{eq:innerprod}
\end{equation}
(where there is an implicit sum over the repeated index $\alpha$), and this is represented diagrammatically as shown in \fref{fig:vector}(iii).

For anyons, as shown in \fref{fig:vector}(iv), we add in a diagrammatic representation of the fusion tree by drawing it on the open end of the index, illustrating the basis whose elements are enumerated by $\alpha$. In this paper we draw the fusion tree in pale grey to indicate that this tree is \emph{not} associated with any diagrammatic factors as per Refs.~\onlinecite{bonderson2007,bonderson2008}. The conjugate of \fref{fig:vector}(iv) is shown in \fref{fig:vector}(v), and is again obtained by vertical reflection of the diagram and complex conjugation of the coefficients $c^\alpha$ to obtain $c^\dagger_\alpha$. Finally, \fref{fig:vector}(vi) is the anyonic version of the inner product, \fref{fig:vector}(iii). To evaluate this diagram, first confirm that the two pale grey fusion trees are mirror images of one another, indicating that $c^\alpha$ and $c^\dagger_\alpha$ are in conjugate bases, and then eliminate them to obtain \fref{fig:vector}(vii) [which is simply \fref{fig:vector}(iii) supplemented by an orientation arrow on the shared index] and thus \Eref{eq:innerprod}. Note that no numerical factor arises from the elimination of the grey fusion trees, as they merely indicate the basis employed on index $\alpha$. Because these fusion trees are not associated with the diagrammatic equations and normalisation factors of the convention described in Refs.~\onlinecite{bonderson2007,bonderson2008}, we call these \emph{implicit} fusion trees, being implied by the choice of basis on $\alpha$. We will contrast this with \emph{explicit} fusion trees, which are associated with the numerical coefficients and rules of the diagrammatic isotopy convention, in \sref{sec:fusesplit}.
 
\subsubsection{Two-index operators and three-index tensors}

\begin{figure}
\begin{center}
\includegraphics[width=\columnwidth]{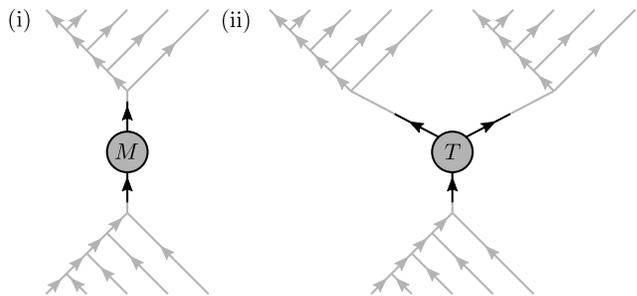}
\caption{Example diagrammatic representations of (i)~an anyonic operator and (ii)~an anyonic three-index tensor (this example has two downgoing indices and one upgoing index; one downgoing index and two upgoing indices is also acceptable). Note that the tensors $M$ and $T$ implicitly play the role of vertices connecting the upper and lower fusion trees, and thus matrix $M^\alpha_\beta$ contains non-zero entries only for values of $\alpha\equiv(a,\mu_a)$ and $\beta\equiv(b,\mu_b)$ such that $a=b$. Similarly, if $\gamma\equiv(c,\mu_c)$ then $T_\gamma^{\alpha\beta}$ is non-zero only where the product $a\times b\rightarrow c$ is permitted by the fusion rules of the anyon model.\label{fig:opsAndTensors}}
\end{center}
\end{figure}

This notation generalises to operators (matrices) and three-index tensors as shown in \fref{fig:opsAndTensors}.
Note that for objects with two or more indices, the total anyonic charge on a fusion tree may take values other than $\mbb{I}$. That is, if an index $\alpha$ is decomposed into a pair $(a,\mu_a)$, then all values of $a$ are allowed provided they are associated with valid labellings of the fusion tree. This is particularly important for operators acting on a subsystem, which may have non-trivial total charge even though the system as a whole carries a total charge of $\mbb{I}$. We find it convenient to adopt a convention where all arrows on fusion trees are upgoing, and the entries of a matrix $M^\alpha_\beta\equiv M^{(a,\mu_a)}_{(b,\mu_b)}$ are then constrained to be non-zero only for $a=b$. For a three-index tensor an entry may be non-zero only where the implicit vertex between $a$, $b$, and $c$ is permitted by the fusion rules. 

\subsubsection{Charge multiplicities}

Considering a three-index tensor $T^\gamma_{\alpha\beta}$ as in \fref{fig:opsAndTensors}, suppose that the fusion rules admit outcomes with multiplicity greater than one, for example
\begin{equation}
8\times 8\rightarrow 1+8+8+\ldots\,.\label{eq:8times8}
\end{equation}
The identical fusion outcomes are commonly distinguished by means of a multiplicity index associated with the fusion vertex, denoted $\mu$. In the graphical representation, however, a three-charge vertex carries only three indices. Consequently, one of the charges iterates not only over the output charges, but also over the associated degeneracy index. Thus, given the fusion rule in \Eref{eq:8times8} and component $a$ of $\alpha$ and component $b$ of $\beta$ satisfy $a=b=8$, then component $c$ of label $\gamma$ takes values in $\{1,8_1,8_2,\ldots\}$ with the subscript denoting the multiplicity index. 

If values in $a$ and $b$ also carry multiplicity indices, then $c$ may carry as many different multiplicity indices as are required to distinguish each unique pair $ab$ and fusion outcome. For example, if $a\in\{8_1,8_2\}$ and $b\in\{8_1,8_2\}$ then $c\in\{1_1,1_2,1_3,1_4,8_1,\ldots,8_8,\ldots\}$. There may come a point, however, at which information about the degeneracy indices associated with the formation of charges $a$ and $b$ is no longer required. At this time, charges $a$ and $b$ may be collapsed such that
\begin{align}
a&\in\{8\},\quad \mu_a=\sum_{i=1}^2 \mu_{a_i}\\
b&\in\{8\},\quad \mu_b=\sum_{i=1}^2 \mu_{b_i},
\end{align}
and similarly,
\begin{equation}
c\in\{8_1,8_2\},\quad \mu_{c_1}=\sum_{i~\mrm{odd}} \mu_{c_i},\quad \mu_{c_2}=\sum_{i~\mrm{even}} \mu_{c_i}
\end{equation}
where it has been assumed that odd $i$ in $\mu_{c_i}$ are associated with the first 8 in \Eref{eq:8times8} and even $i$ are associated with the second.

\subsection{Manipulation\label{sec:manipulation}}

In this Section we present all single-tensor operations required for the implementation of anyonic DMRG. These include hermitian conjugation of multi-index objects, fusing and splitting of indices, conversion between upper and lower indices using caps and cups, and changes of fusion tree basis using braiding and $F$-moves. 

\subsubsection{Hermitian conjugation}

The Hermitian conjugation of anyonic vectors in the graphical notation was introduced in \Eref{eq:hermconj} and \fref{fig:vector}(v). Extension to multiple-index objects (for example $T^{\alpha\beta}_\gamma$) is straightforward: All entries in the object are replaced by their complex conjugates, while all upper indices become lower indices and vice versa while retaining their left-to-right ordering, e.g.
\begin{equation}
(T^\dagger)^\gamma_{\alpha\beta} = \left(T^{\alpha\beta}_\gamma\right)^*\quad\forall~\alpha,\beta,\gamma,
\end{equation}
and the associated graphical representation is reflected vertically.

\subsubsection{F-moves}

The simplest manipulation of a single tensor with which we will be concerned is the $F$-move, which is a change of basis on a composite index $\alpha$ corresponding to an alternative choice of implicit fusion tree. The $F$-move is a unitary transformation which may be applied to any part of a unidirectional fusion tree (one made entirely out of the same orientation of vertex) according to the rule
\begin{equation}
\raisebox{-28pt}{\includegraphics[width=3in]{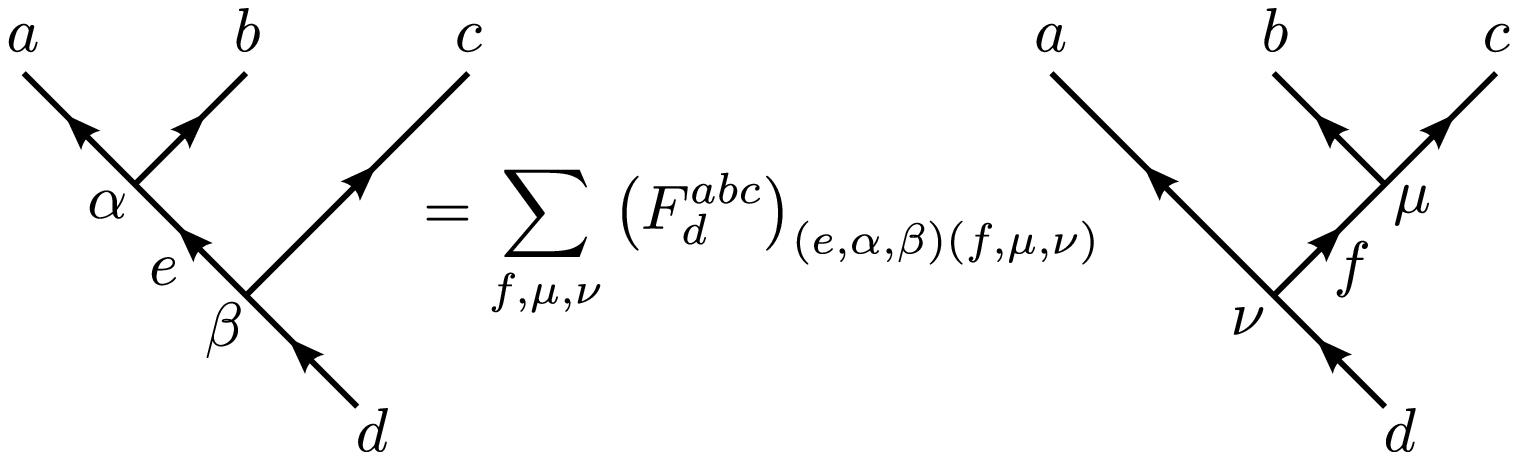}}
\end{equation}
where the ten-index tensor $[F^{abc}_d]_{(e\alpha\beta)(f\mu\nu)}$ is specified by the anyon model. In this paper we use only $F$-moves performed at the trunk of the fusion tree, though they may be freely performed at any level of the tree, on implicit or explicit indices. Note that for more extensive fusion trees, such as that shown in Figures~\ref{fig:intro}(iii) and~\ref{fig:vector}(iv), it is common for multiple values of $\alpha$ to transform identically. In this example, if performing a transformation at the trunk of the fusion tree, the $F$-coefficients are insensitive to the labels
$a_1$, $a_2$, $a_3$, and $a_7$,
and it may be advantageous to treat these indices collectively as if they corresponded to a degeneracy of charge $a_8$. This is particularly true if the leaves of the fusion tree are supplemented with auxiliary degrees of freedom such as position data, or accessory qubits, as these may always be %
treated as degeneracies of the corresponding charge label, and thus all acquire the same multiplicative coefficients from operations on the fusion tree, consistently across multiple anyonic manipulations.

For inverse $F$-moves the relevant coefficients may be obtained by recognising that for fixed $a,b,c,d$ the $F$ tensor may be written as a matrix mapping between the initial and final bases, and this matrix is invertible,
\begin{equation}
\begin{split}
\sum_{(f,\mu,\nu)}&\left[F^{abc}_d\right]^{-1}_{(e,\alpha,\beta)(f,\mu,\nu)} \left[F^{abc}_d\right]_{(f,\mu,\nu)(g,\rho,\sigma)} \\
&= \delta_{(e,\alpha,\beta)(g,\rho,\sigma)}=\delta_{eg}\delta_{\alpha\rho}\delta_{\beta\sigma}.
\end{split}
\end{equation} 
For $F$-moves on the trees associated with kets, the coefficients may be obtained by vertical reflection, arrow reversal (which takes the place of conjugation of charges), and complex conjugation of the coefficients of $F$ or $F^{-1}$ as appropriate.

\subsubsection{Combining and splitting indices\label{sec:fusesplit}}

\paragraph{Index fusion and splitting operations:\label{sec:fusesplitoperations}}

\begin{figure}
\includegraphics[width=\columnwidth]{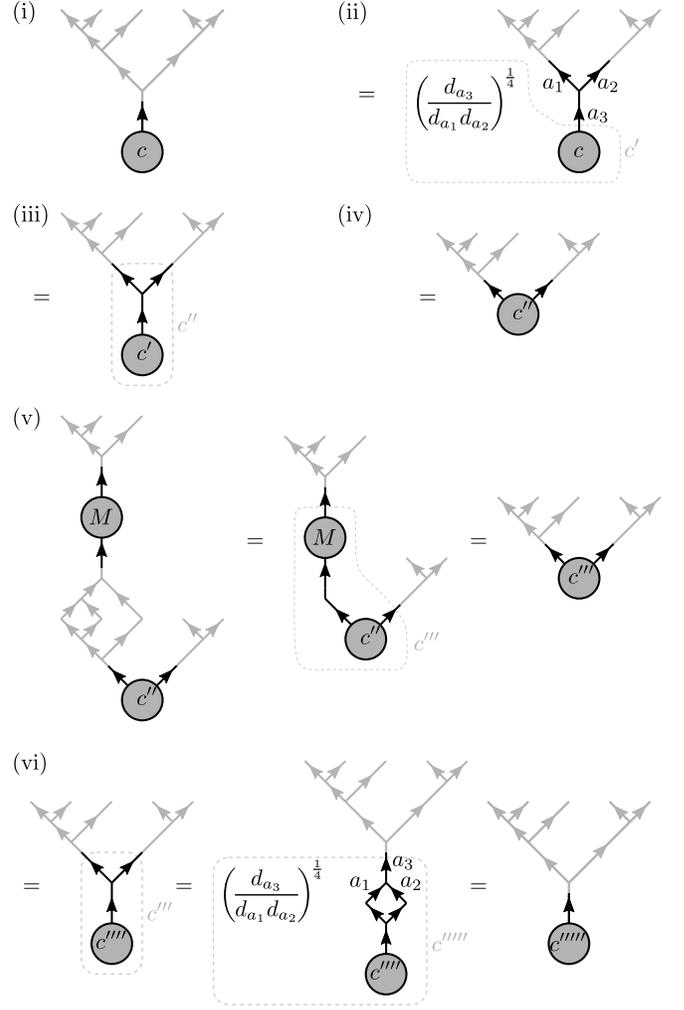}
\caption{Splitting and fusing of indices: (i)~One-index representation of a state of five anyons on the disc. The fusion tree basis has been chosen for subsequent convenience. (ii)~The trunk vertex is made explicit, and (iii)~normalisation factors are absorbed into a redefinition of the vector, $c^\alpha\rightarrow c'^\alpha$. (iv)~The vertex is then absorbed to yield a two-index representation of the state. (v)~Application of a three-site operator $M$ to state $c$. (vi)~Application of a normalised fusion vertex to the greek indices combines two greek indices back into one.\label{fig:splitting}}
\end{figure}

Consider again a vector $c^\alpha$ describing the state of five anyons on the disc, as per \fref{fig:vector}(iv). 
Suppose there exists an operator $M^\beta_\gamma$ acting on only three consecutive anyons, which we wish to apply to $a_1$, $a_2$, and $a_3$. We proceed as follows: First, we use $F$-moves to place $c^\alpha$ into a more convenient basis as shown in \fref{fig:splitting}(i). 
We then introduce an explicit vertex normalised in accordance with the diagrammatic isotopy convention [\fref{fig:splitting}(ii)]. Unlike the implicit vertices used so far, which merely provide a mnemonic for the basis in use on a given index, this vertex is accompanied by a normalisation factor and must be manipulated in accordance with the rules specified in \bonderson{}. We then define the two-index tensor $c^{(3)\beta\gamma}$ of \fref{fig:splitting}(iv) by absorbing both the associated numerical coefficients and the vertex into $c^\alpha$ [Figs.~\ref{fig:splitting}(iii)-(iv)]. Because the charges on the vertex absorbed into $c''^{\beta\gamma}$ may always be uniquely determined from those appearing on the remaining implicit trees, there is a 1:1 mapping between values of $\alpha$ and pairs $\beta\gamma$, and if we associate
\begin{align*}
\alpha&\equiv (a,\mu_a)\\
\beta&\equiv (b,\mu_b)\\
\gamma&\equiv (c,\mu_c)
\end{align*}
then for a given correspondence between $\alpha$ and a pair $\beta\gamma$,
\begin{equation}
c''^{\beta\gamma}=\left(\frac{d_a}{d_bd_c}\right)^\frac{1}{4} c^\alpha.
\end{equation}
Noting that $c^\alpha$ is a state, we have $a=\mbb{I}$ and the above expression simplifies by identifying $d_a=1$, $b=\overline{c}$, and thus $d_b=d_c$. Operator $M$ may now be applied to the state as shown in \fref{fig:splitting}(v), yielding the new state
\begin{equation}
c'''^{\beta\gamma} = M^\beta_\delta c''^{\delta\gamma}.
\end{equation}
The indices appearing on $c'''$ may then be recombined. Formally, the absorbed vertex is re-expressed as per \fref{fig:splitting}(iii) and is then surmounted by a normalised fusion vertex, as per \fref{fig:splitting}(vi). The loop is evaluated according to the rule
\begin{equation}
\raisebox{-20pt}{\includegraphics[width=0.5in]{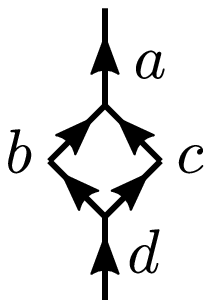}}=\,\delta_{ad}\sqrt{\frac{d_{b}d_{c}}{d_{a}}},\label{eq:looprule}
\end{equation}
and these factors are absorbed into the redefined vector $(c''''')^\alpha$ where they cancel out the vertex normalisation factors of $[d_{a_3}/(d_{a_1}d_{a_2})]^{1/4}$ from Figs.~\ref{fig:splitting}(ii) and~(vi). %
In practice, the fusion operation taking $c'''$ into $c'''''$ is always equivalent to reversing the splitting process.

The index-splitting process may also be applied to two-index objects to yield three-index objects. It is customary not to produce trivalent vertices having three upward or three downward legs, as such objects may always be rewritten in terms of vertices having at most two legs in the same direction and either an additional identity charge or a bend. In this paper we extend the same convention to the anyonic tensors we employ. We also do not create objects with four or more indices, as the labelling of the internal vertices becomes ambiguous for non-Abelian anyonic models. Such objects are anyway unnecessary for the implementation of any tensor network algorithm, as an $n$-index conventional tensor may always be replaced by an anyonic tensor having three or less indices, but $n$ leaves, as we shall see in \sref{sec:networks}.

The index-fusing process may also be applied to three-index tensors to obtain two-index tensors, though we note that any time a pair of indices are to be combined, either both must be upgoing indices, or both must be downgoing indices, to permit their connection with a fusion vertex as per \fref{fig:splitting}(v). Upgoing and downgoing indices may be combined, but to do so it is first necessary to convert both into the same type using a bend, as discussed in \sref{sec:bending}.

From here on, to reduce the use of primes we will use the same label (e.g.~$c$) to represent an object both before and after a single-tensor operation. Thus, for example, in \fref{fig:splitting} the objects $c$, $c'$, and $c''$ would all be denoted $c$.

\paragraph{Mixed normalisation:\label{sec:notenorm}}

The reader might wonder why, in \sref{sec:fusesplitoperations}, we have chosen to combine two different normalisation schemes, namely the implicit fusion tree normalisation scheme, in which no factors arise from the fusion tree diagrams, and the diagrammatic isotopy normalisation scheme used in \bonderson{}. The answer is that in doing so, we obtain the best of both worlds: Our vectors, matrices, and three-index tensors exhibit a natural normalisation where states satisfy $c^\alpha c^\dagger_\alpha = 1$, the identity operator is described by the identity matrix, and index contractions may be performed without having to evaluate any diagrammatic factors provided the implicit trees match up. 
However, by switching to the diagrammatic isotopy basis on all vertices absorbed into our tensors, if we now connect these tensors into a network which permits elimination of the (pale grey) implicit fusion trees, then after eliminating these trees we may manipulate this network using diagrammatic isotopy. The same is also true for any subnetwork normalised entirely within the diagrammatic isotopy convention. %

Another advantage to using the diagrammatic normalisation convention for vertices absorbed into our tensors is that we may use vertical bends to convert upper greek indices into lower indices, and vice versa. We may include vertical bends in our tensor networks, and we may introduce cup/cap pairs on greek indices at will, which is extremely useful when contracting networks. The implementation of vertical bends is described in \sref{sec:bending}, and their use in tensor networks is discussed in \sref{sec:networks}.

For consistency, operations should only ever be performed on regions of a network which all adhere to the same normalisation convention. Thus, for example, both vertices acted on by an $F$-move must be normalised according to the implicit tree convention, or both must be normalised according to the diagrammatic isotopy convention. 
Also note that the implicit vertex normalisation scheme only has meaning in the context of a greek index, where the labellings of the implicit fusion tree are enumerated by that index. Consequently, the only operation where we can change the normalisation of a vertex is during the fusing or splitting of a greek index.

\paragraph{Using fusion and splitting tensors:}

In \sref{sec:fusesplitoperations} we found it convenient to describe splitting and fusing in terms of operations performed on a tensor such as $c^\alpha$ or $M^\alpha_\beta$, in which context it may be considered a single-tensor operation. However, this process may also be understood as contraction with a three-index fusing or splitting tensor [with the latter also subsuming factors arising from the loop diagram in \fref{fig:splitting}(v)]. Defining
\begin{align}
\left(\tilde{N}_\mrm{split}\right)^{\alpha\beta}_\gamma &= \delta^c_{ab} \left(\frac{d_a}{d_bd_c}\right)^\frac{1}{4}\\
\left(\tilde{N}_\mrm{fuse}\right)_{\alpha\beta}^\gamma &= \delta^c_{ab} \left(\frac{d_bd_c}{d_a}\right)^\frac{1}{4}
\end{align}
we may write
\begin{equation}
c^{\alpha\beta}=\left(\tilde{N}_\mrm{split}\right)^{\alpha\beta}_\gamma c^\gamma \qquad
c^\alpha = \left(\tilde{N}_\mrm{fuse}\right)_{\beta\gamma}^\alpha c^{\beta\gamma}
\end{equation}
where $N^c_{ab}$ specifies the anyonic fusion rules
\begin{equation}
a\times b \longrightarrow \sum_c N^c_{ab} c
\end{equation}
and $\delta^c_{ab}$ is 1 if $N^c_{ab}$ is non-zero and 0 otherwise. Thus $\tilde{N}_\mrm{fuse}$ and $\tilde{N}_\mrm{split}$ are analogous to $\Upsilon_\mrm{fuse}$ and $\Upsilon_\mrm{split}$ in \rcite{singh2011}, and $\tilde{N}_\mrm{split}$ corresponds to $\tilde{N}$ in Refs.~\onlinecite{pfeifer2010} and \onlinecite{singh2014}. 

We note that our definition of $\tilde{N}_\mrm{fuse}$ differs from that in Refs.~\onlinecite{pfeifer2010} and \onlinecite{singh2014}, where $\tilde{N}_\mrm{fuse}=\tilde{N}_\mrm{split}=\tilde{N}$, as we are employing a tensor-by-tensor approach in the present paper, and thus the factors $(d_bd_c/d_a)^{1/2}$ arising from the loop in \fref{fig:splitting}(v) are absorbed into state $c$ at the time of index fusion. We therefore find it convenient to incorporate these factors into $\tilde{N}_\mrm{fuse}$. In contrast, in \rcite{singh2014} the numerical factors arising from loop contraction are evaluated for the tensor network as a whole, rather than on an operation-by-operation basis, and thus for fusing as well as splitting, only the normalisation factors $[d_a/(d_bd_c)]^{1/4}$ are incorporated into $\tilde{N}$. (In \rcite{pfeifer2010} it is noted that the factors associated with the diagrammatic isotopy convention must be accounted for separately, though no explicit recipe was given for doing so. Here we give a practical recipe for doing so without having to separately evaluate the numeric factors associated with labellings of a tensor network, though this comes at the cost of losing the equivalence of $\tilde{N}_\mrm{fuse}$ and $\tilde{N}_\mrm{split}$.)

\paragraph{Block structure and fusion priority:\label{sec:fusionpriority}}

Although we will not be using the fusing and splitting tensor formalism in this paper, there is one important thing which we can learn from them. First, consider that each greek index $\alpha$ represents a pair of indices $(a,\mu_a)$ corresponding to charge and degeneracy respectively. As mentioned in \sref{sec:tensorconstruction}, when taken in conjunction with the anyonic fusion rules, these indices impose a natural block structure on anyonic tensors as the entries of the tensors may be non-zero only where the index charges (supplemented by trivial charges where there are less than three indices on a tensor) are consistent with a valid fusion tree vertex. For example, given an operator $M^\alpha_\beta$ and adhering to the convention that all orientation arrows point upwards, the vertex and the block structure associated with this operator are shown in \tref{tab:Mwithfig}.
\begin{table}[bp]%
\caption{Entries in the matrix representation of an operator $M^\alpha_\beta$ may be non-zero only where the corresponding charge labels yield a valid labelling of a fusion tree vertex. Adhering to the convention that all orientation arrows point upwards, and inserting a third, trivial charge as $M$ has only two indices, we obtain the vertex and block structure associated with $M$. For this example it is assumed that there are only two charges in the anyon model, 0 and~1.\label{tab:Mwithfig}}
Vertex structure: \raisebox{-15pt}{\includegraphics[width=0.5in]{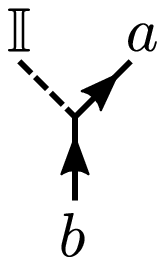}}
\\
Block structure: Entries which may be non-zero are marked with a star: $*$.\\~\\
\begin{tabular}{|c||c|c|c|c|c|c|}
\hline
&$a=0$&$a=0$&$a=0$&$a=1$&$a=1$&$a=1$\\
&$\mu_a=1$&$\mu_a=2$&$\mu_a=\ldots$&$\mu_a=1$&$\mu_a=2$&$\mu_a=\ldots$\\
\hline\hline
$b=0$&\multirow{2}{*}{$*$}&\multirow{2}{*}{$*$}&\multirow{2}{*}{$\cdots$}&&&\\
$\mu_b=1$&&&&&&\\
\hline
$b=0$&\multirow{2}{*}{$*$}&\multirow{2}{*}{$*$}&\multirow{2}{*}{$\cdots$}&&&\\
$\mu_b=2$&&&&&&\\
\hline
$b=0$&\multirow{2}{*}{$\vdots$}&\multirow{2}{*}{$\vdots$}&\multirow{2}{*}{$\ddots$}&&&\\
$\mu_b=\ldots$&&&&&&\\
\hline
$b=1$&&&&\multirow{2}{*}{$*$}&\multirow{2}{*}{$*$}&\multirow{2}{*}{$\cdots$}\\
$\mu_b=1$&&&&&&\\
\hline
$b=1$&&&&\multirow{2}{*}{$*$}&\multirow{2}{*}{$*$}&\multirow{2}{*}{$\cdots$
}\\
$\mu_b=2$&&&&&&\\
\hline
$b=1$&&&&\multirow{2}{*}{$\vdots$}&\multirow{2}{*}{$\vdots$}&\multirow{2}{*}{$\ddots$}\\
$\mu_b=\ldots$&&&&&&\\
\hline
\end{tabular}
\end{table}%
When fusing indices, it is desirable to explicitly preserve this block structure. Thus a vector $c^{\beta\gamma}$ will exhibit a block structure with regards to the associated charges $b$ and $c$, having non-zero entries only where $b=\overline{c}$, but the single-index representation $c^\alpha$ should also exhibit block structure with regards to the charge $a$. This is best illustrated with a specific example, so let us posit the $\mbb{Z}_2$ fusion rules with charge labels~0 and~1 satisfying
\begin{equation}
\begin{split}
0\times 0&\rightarrow 0\\
0\times 1&\rightarrow 1\\
1\times 0&\rightarrow 1\\
1\times 1&\rightarrow 0,
\end{split}\label{eq:Z2}
\end{equation}
and $\mu_b$ and $\mu_c$ taking only a value of~1.
An example mapping between pairs $\beta\gamma$ and values of $\alpha$ which preserves the block structure is given in \tref{tab:fusion}.
\begin{table}[bp]%
\caption{Fusion of two indices $\beta$ and $\gamma$ into a single index $\alpha$ requires a mapping between pairs $\beta\gamma$ and values of $\alpha$. It is preferable that this mapping be made in a manner which preserves the block structure of the tensor, such as the example given here.\label{tab:fusion}}
\begin{tabular}{|c|c|}
\hline
$\beta$&$(b,\mu_b)$\\
\hline\hline
1&(0,1)\\
2&(0,2)\\
3&(1,1)\\
4&(1,2)\\
\hline
\multicolumn{2}{c}{~}\\
\hline
$\gamma$&$(c,\mu_c)$\\
\hline\hline
1&(0,1)\\
2&(0,2)\\
3&(1,1)\\
4&(1,2)\\
\hline
\end{tabular}
~~~
\begin{tabular}{|c|c|c|c|c|}
\hline
$\alpha$&$(a,\mu_a)$&$\beta,\gamma$&$(b,\mu_b)$&$(c,\mu_c)$\\
\hline\hline
1&(0,1)&1,1&(0,1)&(0,1)\\
2&(0,2)&1,2&(0,1)&(0,2)\\
3&(0,3)&2,1&(0,2)&(0,1)\\
4&(0,4)&2,2&(0,2)&(0,2)\\
5&(0,5)&3,3&(1,1)&(1,1)\\
6&(0,6)&3,4&(1,1)&(1,2)\\
7&(0,7)&4,3&(1,2)&(1,1)\\
8&(0,8)&4,4&(1,2)&(1,2)\\
9&(1,1)&1,3&(0,1)&(1,1)\\
10&(1,2)&1,4&(0,1)&(1,2)\\
11&(1,3)&2,3&(0,2)&(1,1)\\
12&(1,4)&2,4&(0,2)&(1,2)\\
13&(1,5)&3,1&(1,1)&(0,1)\\
14&(1,6)&3,2&(1,1)&(0,2)\\
15&(1,7)&4,1&(1,2)&(0,1)\\
16&(1,8)&4,2&(1,2)&(0,2)\\
\hline
\end{tabular}
\end{table}%
This particular mapping was constructed as follows:
\begin{enumerate}
\item Let $a$ take the value 0.
\item Select the first fusion rule in \Eref{eq:Z2} consistent with an output charge of $a$.\label{step7}
\item Let $b$ and $c$ be the left and right input charges to this rule respectively.
\item Select the first value of $\beta$ consistent with charge $b$.\label{step5}
\item Select the first value of $\gamma$ consistent with charge $c$.\label{step3}
\item Associate the pair $\beta\gamma$ with the first unassigned value of $\alpha$.\label{step1}
\item Select the next value of $\gamma$ consistent with charge $c$ and return to step~\ref{step1}. If no further compatible values of $\gamma$ are available, advance to step~\ref{step2}.
\item Select the next value of $\beta$ consistent with charge $b$ and return to step~\ref{step3}. If no further compatible values of $\beta$ are available, advance to step~\ref{step4}.\label{step2}
\item Select the next fusion rule in \Eref{eq:Z2} consistent with an output charge of $a$, and return to step~\ref{step5}. If no further compatible rules are available, advance to step~\ref{step6}.\label{step4}
\item Increment the value of $a$. If this results in an invalid value of $a$, stop. Otherwise return to step~\ref{step7}.\label{step6}
\end{enumerate}
In such a construction, we describe $\gamma$ as the \emph{fast-cycling} index and $\beta$ as the \emph{slow-cycling} index.

Now consider the definition of hermitian conjugation given in \sref{sec:tensorconstruction}. Consistency requires that if the fusion of upgoing indices is performed using $(\tilde N_\mrm{fuse})^\alpha_{\beta\gamma}$, then the fusion of downgoing indices is performed using the hermitian conjugate, $(\tilde N^\dagger_\mrm{fuse})_\alpha^{\beta\gamma}$. Consequently, if the rightmost index of a pair is fast-cycling during fusion of upgoing indices, then the rightmost index of a pair must also be fast-cycling during fusion of downgoing indices, and this applies both to charge (e.g.~$b$) and degeneracy (e.g.~$\mu_b$) indices. When fusing indices in software such as \MATLAB{}, it is important to recognise that the software cannot distinguish between upgoing and downgoing indices. Instead, there is a numbered ordering assigned to the indices on a tensor, and in \MATLAB{} when combining a pair of indices the first index is fast-cycling and the second is slow-cycling. Thus it is important that if numbers are assigned to upper indices from left to right, then lower indices must also be numbered from left to right, in order to ensure that fusion is performed in a consistent fashion. One may, of course, number indices in any order if appropriate fast and slow cycling is manually enforced.

Finally, having made the block structure of anyonic tensors explicit in \tref{tab:Mwithfig}, it is useful to point out that the total charge index $a$ and its associated degeneracy index $\mu_a$ may be identified with indices $a$ and $\alpha_a$ in \rcite{singh2010} below Eq.~(4). The third index in this reference, $m_a$, is omitted for anyon models. This is because \rcite{singh2010} discusses the block decomposition of tensors having group symmetries, and for symmetry groups this index enumerates states within an irrep, while for many anyon models no comparable structure exists.

\subsubsection{Vertical bending\label{sec:bending}}

The next operation we will consider is vertical bending of a greek index, converting upper indices into lower indices and vice versa. A recipe for evaluating vertical bends in fusion tree diagrams is given in \rcite{bonderson2007} [though note there is a typo on the lower index of $F$ in Eq.~(2.31)], using the notation of Frobenius--Schur indicators. For simplicity we shall avoid this notation, adhering instead to our convention of labelling the orientations of all line segments with upward-pointing arrows, though our treatment will be equivalent. 

\begin{figure}
\includegraphics[width=\columnwidth]{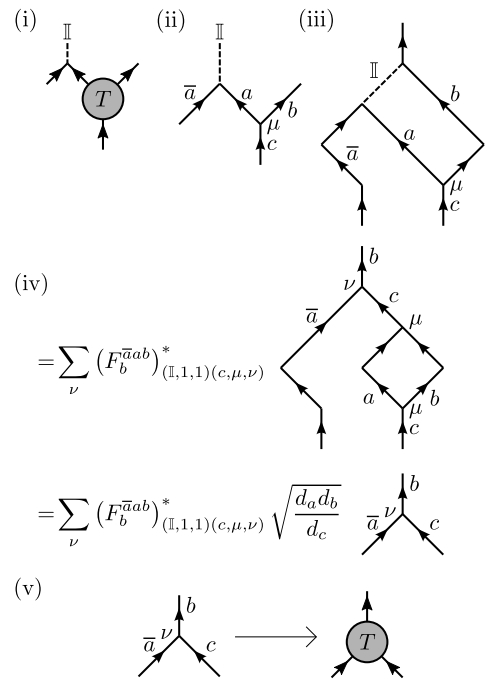} %
\caption{Bending converts a lower index into an upper index. Implicit fusion trees are not shown. (i)~A tensor $T$ is adjacent to a bend vertex. This vertex involves the identity charge and the dual to the charge on the bending leg. (ii)~Consider the fusion diagram associated with each charge labelling of~(i) in turn. (iii)~Since the vertex contained within the tensor is normalised according to the diagrammatic isotopy convention, the legs corresponding to greek labels respect the rules of diagrammatic isotopy. We may therefore deform the diagram as shown, and introduce an additional line carrying the trivial charge. (iv)~We then perform an $F$-move, and contract the resulting loop using \protect{\Eref{eq:looprule}}. (v)~The resulting trivalent vertex yields the structure of the resulting tensor. The numerical coefficients of diagram~(iv) are absorbed into the entries of the tensor. Note that charge $a$ has been mapped into its dual.\label{fig:bend}}
\end{figure}

\paragraph{Evaluation of a bend:}
Since the indices appearing explicitly on a tensor are normalised according to the diagrammatic isotopy convention, if a vertical bend appears \emph{immediately adjacent} to a tensor then we may evaluate it as shown in \fref{fig:bend}.
Indeed, this procedure and its clockwise counterpart serve to define the tensors in \rcite{bonderson2007} denoted $[A^{ab}_c]_{\mu\nu}$ and $[B^{ab}_c]_{\mu\nu}$ according to
\begin{align}
\left[A^{ab}_c\right]_{\mu\nu} &= \sqrt{\frac{d_ad_b}{d_c}}\varkappa_a^*\left[F^{\overline{a}ab}_b\right]^*_{(\mbb{I},1,1)(c,\mu,\nu)}\label{eq:bendA}\\
\left[B^{ab}_c\right]_{\mu\nu} &= \sqrt{\frac{d_ad_b}{d_c}}\left(\left[F^{ab\overline{b}}_a\right]^{-1}\right)^\dagger_{(c,\mu,\nu)(\mbb{I},1,1)}\label{eq:bendB}
\end{align}
where
\begin{equation}
\varkappa_a = \left[F^{a\overline{a}a}_a\right]_{(\mbb{I},1,1)(\mbb{I},1,1)}\, d_a,
\end{equation}
and $\varkappa^*_a$ appears in \Eref{eq:bendA} due to the use of a left-facing Frobenius--Schur indicator when specifying the diagram associated with $[A^{ab}_c]_{\mu\nu}$ in \rcite{bonderson2007}. For unitary anyon models, \Eref{eq:bendB} simplifies to the form given in \rcite{bonderson2007} because for unitary models, $([F^{abc}_d]^{-1})^\dagger=F^{abc}_d$.

Note that, as seen in \fref{fig:bend}(i), a bend may be understood as a vertex with a trivial leg, and this vertex must be normalised according to the diagrammatic isotopy convention and hence is drawn in black. For simplicity, when bending we will generally omit drawing the vertex leg carrying the trivial charge.

\paragraph{Introducing new bends:}
One might hope that the approach given in \fref{fig:bend} would suffice for the evaluation of tensor network diagrams which contain bends. In practice, however, we often need to introduce additional bends when performing tensor contractions, as it is often
necessary to temporarily lower an index then later raise it again. For this we need the identities
\begin{equation}\label{eq:vertbends}
\raisebox{-7pt}{\includegraphics[width=3in]{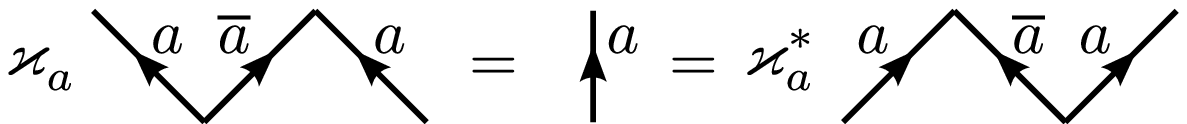}}
\end{equation}
which are valid only if the bending vertices are normalised in the diagrammatic isotopy convention.
For now we concern ourselves primarily with the introduction of bend pairs on an existing greek index. However, in principle we may introduce additional bend pairs anywhere we like in a fusion diagram (or in \sref{sec:networks}, anywhere in an anyonic tensor network), provided we respect the normalisation requirement. This is because the insertion of a bend pair in the middle of an implicit tree has no effect on the number of degrees of freedom involved in the labelling of that tree.

\begin{figure}
\includegraphics[width=\columnwidth]{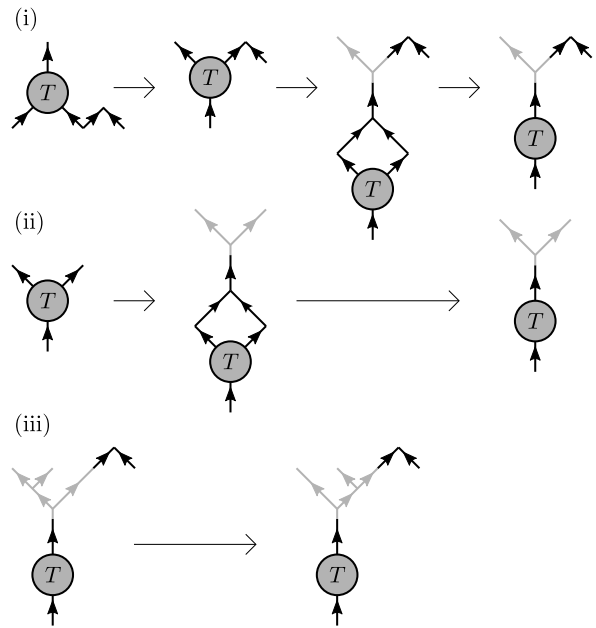}
\caption{(i)~Example of manipulations involving bends performed on an anyonic tensor network. 
(ii)~Restricting our attention just to tensor $T$ in the second step of diagram~(i), the fusion and splitting operations are seen to be entirely routine. The bend does not participate in these operations, and is no more than a deferred instruction to lower the corresponding leaf at a later time, when it is once again present on a tensor as a greek index. (iii)~Similarly, we may perform operations such as $F$-moves and braids so long as all relevant vertices are in a consistent normalisation convention.\label{fig:temporarybend}}
\end{figure}

As mentioned in \sref{sec:notenorm}, we may apply valid operations to any part of a fusion tree diagram provided all vertices within that region share a common normalisation. Thus, if we insert a bend pair (normalised in the diagrammatic isotopy convention) into a portion of the fusion tree which is normalised in the implicit tree convention, we may continue to perform our full range of operations on parts of the tree adjacent to the inserted bends, so long as we do not involve the bending vertices themselves.
For example, if a greek index on a tensor is adjacent to a bend, one may continue to combine that index with others in the normal way as shown in \fref{fig:temporarybend}(i).
In this Figure, two greek indices on tensor $A''$ are fused into a single greek index, and the normalisation convention of the fusion vertex is changed to implicit. Looking exclusively at tensor $A''$ on which the fusion operation acts, this is no different to any other fusion operation, as shown in \fref{fig:temporarybend}(ii). The bend, however, even though it is now separated from tensor $A''$, remains normalised in the diagrammatic isotopy convention at all times. If multiple fusions are performed under a bend, we may even perform $F$-moves on the resulting tree [e.g. \fref{fig:temporarybend}(iii)], because once again, we may restrict our attention only to the portion of the network immediately attached to the tensor on which we are operating, giving a unidirectional 
fusion tree which is normalised entirely in the implicit vertex normalisation convention.
We shall see more worked examples of the insertion of bends when we discuss tensor networks in \sref{sec:networks}.

\paragraph{Bending and index ordering:}
If a greek index carries an implicit fusion tree, %
note that bending this index will cause the branch which was leftmost on the implicit tree before bending to be rightmost after bending, and vice versa. Consequently, if the labellings of this tree are ordered according to \sref{sec:fusionpriority} before bending, then they will not be ordered according to this scheme after bending, differing by a tree-dependent permutation. We avoid explicit calculation of this permutation by recognising that it is always possible to use an appropriate combination of $F$-moves, fusing, and splitting to separate the leaves off one at a time and bend them individually. This
achieves the same transformation while only bending greek indices corresponding to individual leaves, which have no associated implicit fusion tree.

\subsubsection{Braiding}

The final single-tensor manipulation we require is the braiding of two greek indices. As with bending, this follows directly from the equivalent manipulation on a fusion tree. Given a tensor $T^{\alpha\beta}_\gamma$, if we wish to braid index $\alpha$ over index $\beta$ then the entries of $T^{\alpha\beta}_\gamma$ acquire coefficients based on the values of $a$ in $\alpha$, $b$ in $\beta$, and $c$ (and $\mu$, if applicable) in $\gamma$:
\begin{equation}
\raisebox{-40pt}{\includegraphics[width=3in]{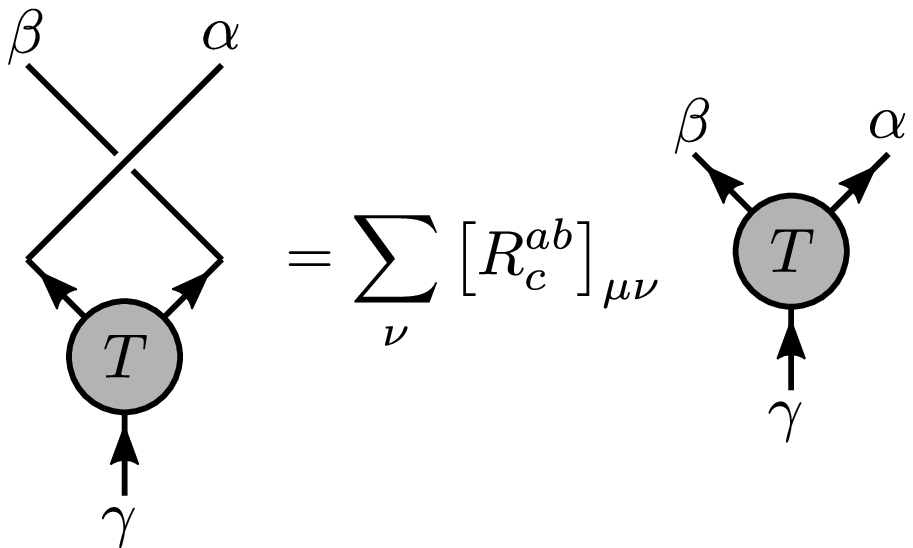}}
\end{equation}
Note that comparison with the equivalent expression for fusion trees,
\begin{equation}
\raisebox{-35pt}{\includegraphics[width=3in]{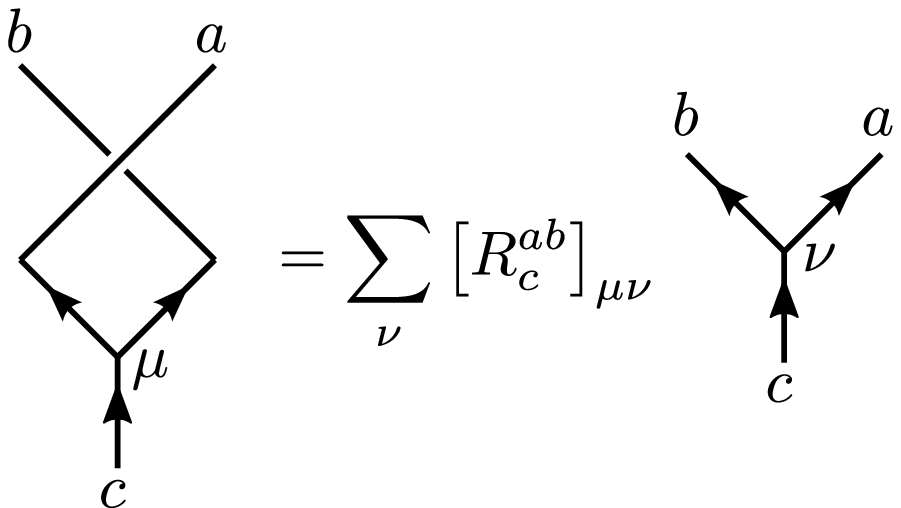}}
\end{equation}
shows that $\mu$ is the degeneracy index associated with the instance of $\gamma$ appearing on the left-hand side of the equation, while $\nu$ is the degeneracy index associated with the instance of $\gamma$ on the right.

Braiding may also be performed within an individual implicit fusion tree, and doing so may occasionally simplify the evaluation of a given tensor network. It is never obligatory, however, and in the interest of keeping our set of basic operations as compact as possible, we shall not concern ourselves with it here.

\subsubsection{$360^\circ$ rotation}

Although this identity is never \emph{necessary}, it can sometimes save computational cost to recognise that when bends are applied to the greek indices of a tensor which are equivalent to performing a full $360^\circ$ rotation of that tensor, this operation is equivalent to the identity.\cite{kitaev2006}

\subsection{Tensor networks\label{sec:networks}}

\subsubsection{Construction\label{sec:networkconstruction}}

\begin{figure}
\includegraphics[width=\columnwidth]{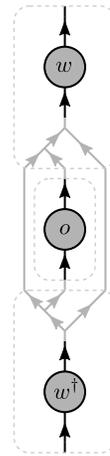}
\caption{A simple example of an anyonic tensor network, made up of three tensors: $w$, $o$, and $w^\dagger$. For illustrative purposes, in this diagram we have grouped tensors with their associated implicit fusion trees using dashed boxes.\label{fig:examplenetwork}}
\end{figure}

Having introduced a graphical notation for anyonic tensors in \sref{sec:tensorconstruction}, we now introduce anyonic tensor networks. To construct an anyonic tensor network, the diagrammatic representations of two or more anyonic tensors are drawn with some leaves of their fusion trees connected. Vertical bends may be included, normalised according to the diagrammatic isotopy convention. Aside from tensors and vertical bends, the only other vertices appearing should be those appearing on the unidirectional fusion trees associated with the greek indices of the anyonic tensors. %
Note that it is not necessary to connect all leaves associated with a greek index on a tensor, and leaves associated with a greek index may be connected to any number of other tensors. An example of an anyonic tensor network is shown in \fref{fig:examplenetwork}.

\subsubsection{Normalisation and diagrammatic isotopy\label{sec:diagisotopy}}

In \sref{sec:fusesplit} we adopted a normalisation scheme combining two different vertex normalisations: ``Implicit'' vertices are drawn in grey and their contraction yields no numerical factors, giving rise to convenient tensor identities such as $c^\alpha c^\dagger_\alpha=1$ for a state vector. ``Explicit'' vertices, such as those absorbed into tensors during fusing and splitting operations, are drawn in black and are associated with the factors of the diagrammatic isotopy convention.\cite{bonderson2007} Many operations behave equivalently in both normalisations, in particular
\begin{itemize}
\item $F$-moves on unidirectional fusion trees (all fusion vertices or all splitting vertices) with both vertices in the same normalisation convention,
\item horizontal deformation of world lines,
\item vertical sliding of diagram portions where this does not cause overlap or bending (this operation is trivial), and
\item braiding.
\end{itemize}
Although we cautioned earlier against applying operations to mixed fusion trees, we now note that because of this equivalence, horizontal deformation, braiding, and vertical sliding operations may be applied to \emph{any} part of a network, regardless of whether normalisation conventions are consistent throughout. By combining the horizontal deformation of world lines with braiding, we may also horizontally push world lines across tensors.
(Strictly speaking, $F$-moves can also be applied to mixed trees, but the bookkeeping in the vicinity of bends can be confusing, so we caution against this.)

Regarding operations which differ between the two conventions:
\begin{itemize}
\item The elimination of loops is formally different but yields equivalent numerical results, because converting vertices from implicit to explicit normalisation introduces numerical factors which exactly cancel out those arising from elimination of the loop in the diagrammatic isotopy convention. 
\item The insertion of additional vertices emitting or absorbing the trivial charge is not supported in the implicit tree convention, as the enumerated labellings do not include this trivial charge.
\item The ability to insert, delete, and evaluate (absorb) vertical bends also differs significantly between the two conventions.
\end{itemize}
In particular, in the implicit tree convention, inconsistencies may arise if an attempt is made to introduce bends, connect them to the rest of the fusion tree through vertices involving trivial charges, and then manipulate these vertices using $F$-moves.
Consequently we only introduce bends in the diagrammatic isotopy convention, only rotate tensors in the diagrammatic isotopy convention, and---for simplicity's sake---never perform operations on fusion trees which involve more than one normalisation convention. %
We can, however, continue to apply the full range of diagrammatic isotopy operations to \emph{sub}diagrams which are entirely normalised in the diagrammatic isotopy convention.

Now let us consider the operations available to us from the perspective of their actions on the greek indices of an individual tensor, with $F$-moves being considered as a change of basis on a single greek index. We see that we may:
\begin{itemize}
\item fuse and split indices,
\item use $F$-moves at trunk level (and also deeper in the fusion tree, though this is never actually necessary) to affect the manner in which combined indices separate on splitting,
\item use horizontal deformations and braiding to re-order greek indices, and
\item use bends to raise and lower greek indices,
\end{itemize}
all within a consistent normalisation scheme. In fact, when applied in close proximity to the tensor like this, all of these operations act on vertices normalised according to the diagrammatic isotopy scheme with the exception of the trunk-level $F$-move, which acts on implicit trees.

Applied appropriately, and supplemented by the pairwise contraction operation discussed in \srefs{sec:tensorconstruction}{sec:pairwise}, this dictionary of fundamental operations is sufficient to contract any anyonic tensor network satisfying the prescriptions of \sref{sec:networkconstruction}, and we need never evaluate operations acting on more remote parts of a tensor's fusion tree.

\subsubsection{Pairwise contraction\label{sec:pairwise}}

Any pair of tensors may be contracted together into a single tensor. In the following, we show how contractions are perfomed by means of illustrative examples.

\begin{figure}
\includegraphics[width=\columnwidth]{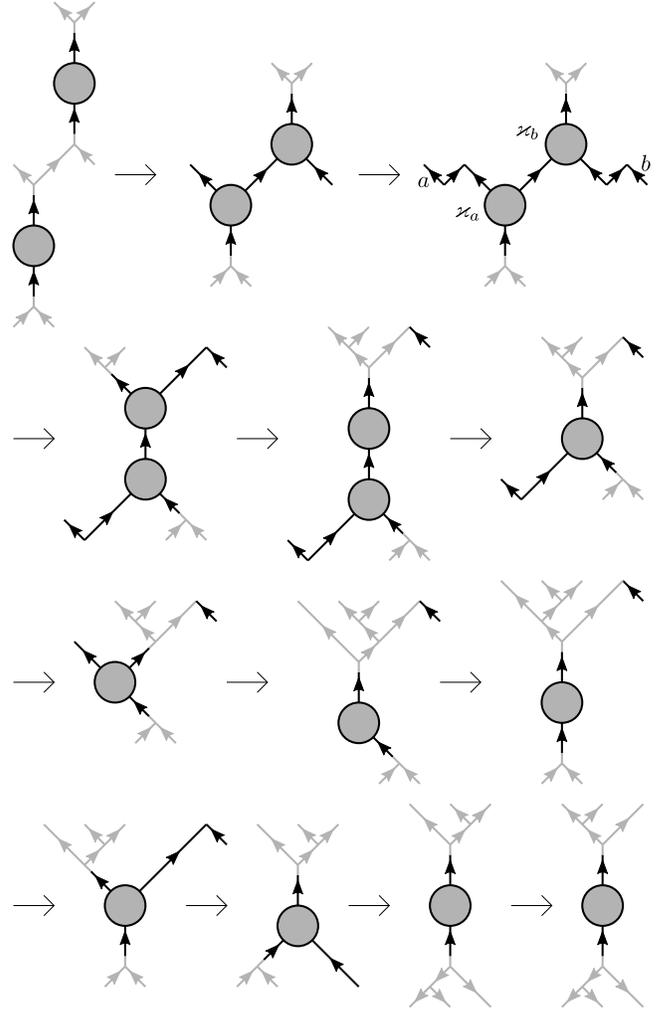} %
\caption{Example contraction procedure for a pair of tensors connected by one leaf. Factors of $\varkappa_a$ and $\varkappa_b$ are written next to the tensors into which they are subsequently absorbed.\label{fig:paircont1}}
\end{figure}

\paragraph{First example: Simple connected pair.} Consider the network given in \fref{fig:paircont1} where two tensors are connected by their leaves.
This example is slightly more complex than the contractions considered in \sref{sec:tensorconstruction}, as the tensors in \fref{fig:paircont1} are connected only by leaves, rather than by greek indices. However, using the single-tensor manipulations of \sref{sec:manipulation}, these tensors may be put into a form where all summed leaves (and only the summed leaves) appear on a single greek index on each tensor, and there are no more than three unsummed greek indices in total. The contraction may then be implemented by means of a summation over that greek index, with the resulting object being a valid anyonic tensor itself having no more than three greek indices. The resulting object may then be reshaped into any desired form%
. %

\begin{figure}
\includegraphics[width=\columnwidth]{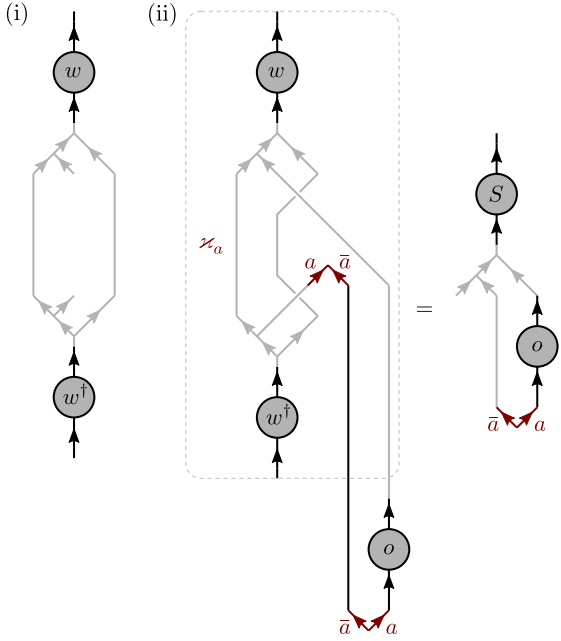}
\caption{COLOUR ONLINE: (i)~Na\"\i{}eve construction of an anyonic superoperator from $w$ and $w^\dagger$ of \protect{\fref{fig:examplenetwork}}. (ii)~Alternative construction of the superoperator as a convex tensor $S$. The action of $S$ on an operator $o$ is as shown. In this Figure and in \pfref{fig:superop2}, where cups, caps, and factors $\varkappa$ originate in an application of \pEref{eq:vertbends} we have coloured them in sets to show which go together, and assigned labels to the charges on the bending indices. The bending vertices continue to be normalised according to the diagrammatic isotopy convention. Where we choose to absorb the cup into one tensor and the cap into another, we may freely choose whether to absorb the corresponding factors $\varkappa$ along with the cup or the cap. In these examples we have chosen to absorb $\varkappa$ with the cap.\label{fig:superop}}
\end{figure}

\begin{figure}
\includegraphics[width=\columnwidth]{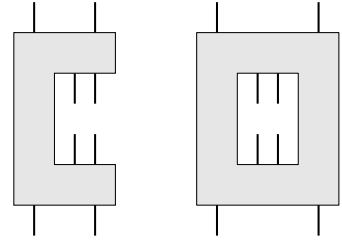}
\caption{Examples of tensors which are not convex tensors.\label{fig:concavetensors}}
\end{figure}

\paragraph{Second example: Non-convex network.} Next, consider the example tensor network given earlier in \fref{fig:examplenetwork}.
In this network, we have three tensors: $w$, $w^\dagger$, and $o$. Suppose that we know $w$ and $w^\dagger$, and wish to find a tensor $o$ such that 
\begin{equation}
wow^\dagger\propto o. \label{eq:superop}
\end{equation}
We could trial different operators for $o$ and use a Lanczos-type 
algorithm to obtain those which satisfy \Eref{eq:superop}, first contracting $o$ with $w$ and then the resulting object with $w^\dagger$, or vice versa, and this would pose no greater challenge than did \fref{fig:paircont1}. Alternatively, however, we might wish to contract $w$ with $w^\dagger$ to obtain a superoperator, write this in matrix form, compute its eigenvectors, and then map these eigenvectors back to operators satisfying \Eref{eq:superop}. If we take this approach, though, then we encounter a complication. The na\"\i{}eve construction of the superoperator, given in \fref{fig:superop}(i), does not seem to correspond to any sort of anyonic tensor we have discussed so far.
This is because our formalism supports only \emph{convex} tensors. A convex tensor is one where
\begin{enumerate}
\item a bounding box may be drawn which completely encloses the fusion tree of the tensor without intersecting it,
\item this bounding box touches the free ends of all leaves of the (unidirectional) fusion trees, and 
\item all upgoing leaves are consecutive on this boundary and all downgoing leaves are consecutive on this boundary.
\end{enumerate}
Thus tensors having forms such as those shown in \fref{fig:concavetensors} are excluded.
Nevertheless, it is still possible to proceed with this calculation. Using the permitted operations described in \sref{sec:diagisotopy} we may deform this diagram
into the form of \fref{fig:superop}(ii), %
and an operator $o$ which is an eigenoperator of \fref{fig:superop}(i) will also be an eigenoperator of superoperator $S$ in \fref{fig:superop}(ii), where $S$ is defined, and acts on $o$, as shown.

\begin{figure}
\includegraphics[width=\columnwidth]{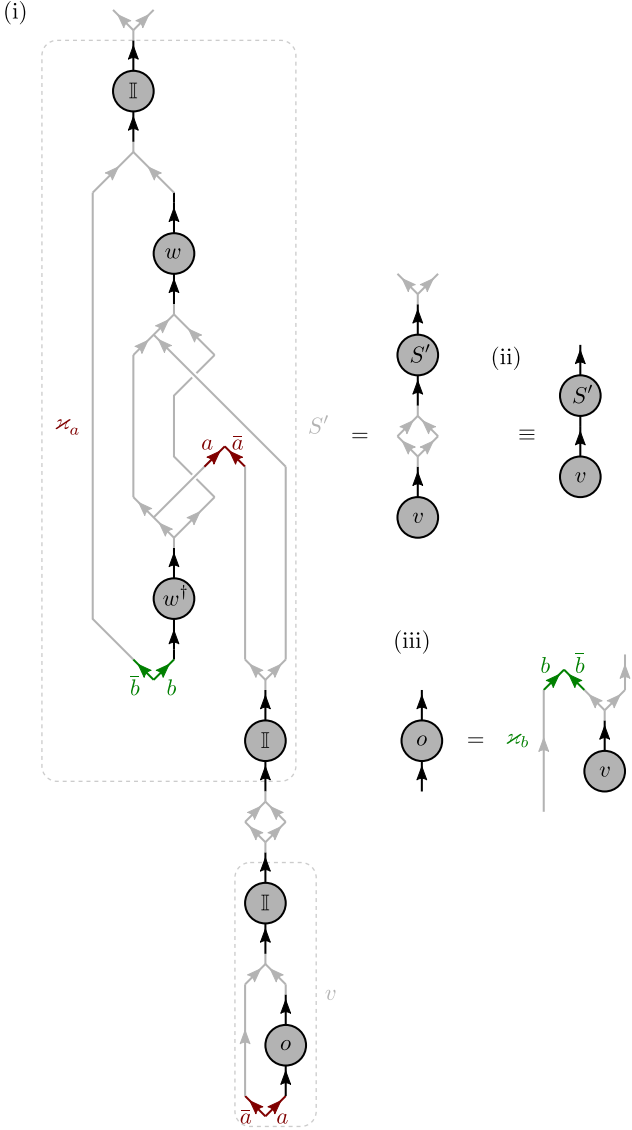}
\caption{COLOUR ONLINE: (i)~The superoperator may also be written as a matrix $S'$. If the operator $o$ is rewritten as a vector $v$ as shown, then the action of $S'$ on $v$ is simply matrix multiplication, $S'v$, as shown in diagram~(ii). Eigenoperators $o$ may be recovered from eigenvectors $v$ as shown in diagram~(iii).\label{fig:superop2}}
\end{figure}

A more satisfying though slightly more complicated approach is shown in \fref{fig:superop2}(i). 
The tensors labelled $\mbb{I}$ are identity operators, which may be freely inserted into a tensor network at any time%
, even when (as here) they end up separating an inserted bend pair.

To understand why we have inserted these identity operators, let us examine the topmost operator more closely. Its lower vertex serves to fuse together two indices during the construction of $S'$. This then necessarily yields a single greek index. The body of the identity operator, $\mbb{I}^\alpha_\beta$, is then absorbed onto this greek index to yield the upper greek index of $S'$ (though in practice this step is trivial as $\mbb{I}^\alpha_\beta$ is just the identity matrix). Finally, the upper vertex of the identity operator forms part of the implicit fusion tree of $S'$. The insertion of this identity operator therefore essentially serves as a prescription as to how the upgoing leaves of the network involving $w$ and $w^\dagger$ are to be fused in order to obtain the upper greek index of $S'$, and what implicit fusion tree is to be associated with this index.

The other two identity operators behave similarly, with one constructing the lower implicit fusion tree of $S'$, and the other the implicit fusion tree of $v$. As can be seen from \fref{fig:superop2}(ii), the eigenvectors $v^\alpha$ corresponding to eigenoperators $o$ may then be obtained simply by diagonalising $S'^\alpha_\beta$, with operators $o^\alpha_\beta$ recovered from $v^\alpha$ as per \fref{fig:superop2}(iii). 
This example %
illustrates a 
useful 
general principle: The anyonic tensor formalism presented here and in \rcite{pfeifer2010} supports only convex tensors, but any anyonic tensor calculation can always be re-expressed in terms of convex tensors, so there is no loss of generality.

\begin{figure}
\includegraphics[width=\columnwidth]{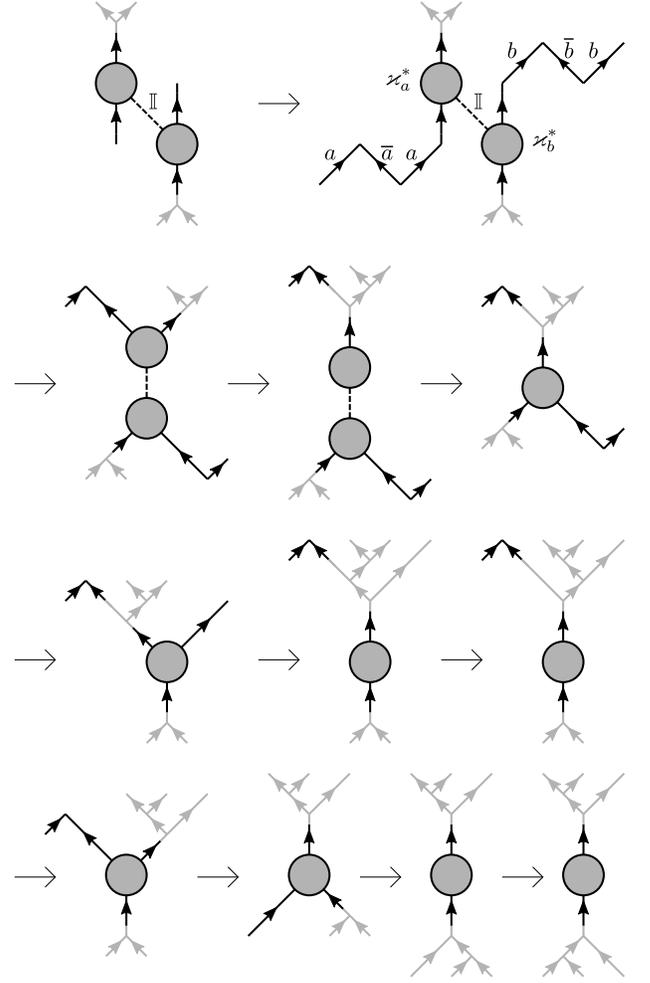}
\caption{Example of contraction of a pair of disjoint tensors.\label{fig:disjoint}}
\end{figure}

\paragraph{Third example: Disjoint tensors.} Finally, we note that where a pair of tensors are not connected by any leaves, pairwise contraction is achieved by fusing greek indices if necessary, so that there are a maximum of two greek indices per tensor, inserting an additional greek index carrying the trivial charge with no degeneracy onto each tensor, upgoing on one tensor and downgoing on the other, and then contracting over this index. The indices should be inserted such that the resulting tensor will be convex. An example is shown in \fref{fig:disjoint}.

\paragraph{Caveat:\label{eq:summationcaveat}} It is important to note that where pairs of anyonic tensors are contracted by index summation, this summation has so far always been over a \emph{single} greek index. When summation is performed over a single greek index, the associated fusion tree is free of loops, and so no extra numerical factors arise from the diagrammatic isotopy convention. In contrast, summing over two greek indices at once does create a diagrammatic loop, and thus (for example) if $A^{\alpha\beta}$ is obtained by splitting index $\gamma$ on $A^\gamma$ and $B_{\alpha\beta}$ is obtained by splitting index $\gamma$ on $B_\gamma$, then
\begin{equation}
A^\gamma B_\gamma = \sum_{c=\mbb{I},~\mu_c} A^{(c,\mu_c)} B_{(c,\mu_c)}
\end{equation}
but
\begin{equation}
A^{\alpha\beta} B_{\alpha\beta} = \!\!\sum_{\substack{a,~\mu_a\\b=\overline{a},~\mu_b}} \!\!A^{(a,\mu_a)(b,\mu_b)} B_{(a,\mu_a)(b,\mu_b)}\,d_a
\end{equation}
where the extra factor of $d_a$ arises from the loop. For simplicity we therefore advocate always fusing greek indices before summation.

\subsubsection{Preferential nature of pairwise contraction}

In Appendix~B of \rcite{pfeifer2014a} it was shown that any network of ordinary tensors may always be optimally contracted by means of a sequence of pairwise contractions. We now argue that the same holds true for anyonic tensor networks.

First, consider that any anyonic tensor admits a decomposition into labelled fusion tree and degeneracy components. In analogy to Eq.~(4) of \rcite{singh2010} we may write
\begin{equation}
\mbb{V} \cong \bigoplus_{a,b,c} \left(\mbb{D}^{ab}_{c}\otimes \raisebox{-14pt}{\includegraphics[width=0.324in]{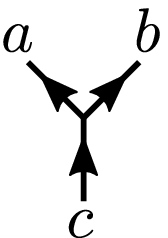}}\,
\right)
\end{equation}
where $\mbb{V}$ is the space of anyonic tensors, $\mbb{D}^{ab}_c$ is the degeneracy space associated with charges $a$, $b$, and $c$, and has dimension $\max{(\mu_a)}\times\max{(\mu_b)}\times\max{(\mu_c)}$ for a given triplet $(a,b,c)$. An anyonic tensor $T^{\alpha\beta}_\gamma$ then admits the decomposition
\begin{equation}
T^{\alpha\beta}_\gamma = \left(D^{ab}_c\right)^{\mu_a\mu_b}_{\mu_c} \otimes \raisebox{-14pt}{\includegraphics[width=0.324in]{vertexabc}}
\end{equation}
where $\left(D^{ab}_c\right)^{\mu_a\mu_b}_{\mu_c}$ for fixed $(a,b,c)$ is the degeneracy tensor associated with a given charge sector of $T^{\alpha\beta}_\gamma$, with entries enumerated by the values of $\mu_a$, $\mu_b$, and $\mu_c$.
Extending this decomposition to the entire tensor network then yields a space of labelled fusion diagrams and, for each labelling, an associated network of degeneracy tensors%
.

\begin{figure}
\includegraphics[width=\columnwidth]{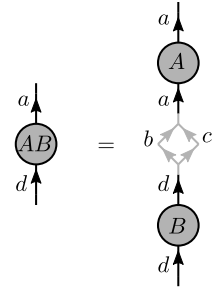} %
\caption{This contraction provides an example of the importance of eliminating loops at the first available opportunity during tensor contraction (see text), and by implication, favouring pairwise contraction of both the degeneracy and the fusion tree components of tensors during the evaluation of a tensor network.\label{fig:paircont2}}
\end{figure}

A relatively popular approach to the contraction of an anyonic tensor network (which we ourselves used in \rcite{singh2014}) is to iterate over all valid labellings of the fusion tree diagram and, for each labelling, contract the associated network of degeneracy tensors using a series of pairwise contractions, then multiply by the factor associated with the labelled fusion tree diagram for the entire network. This approach is conceptually simple, but is, in general, suboptimal (though for the TEBD algorithm considered in \rcite{singh2014} the overhead is relatively small, being at most a factor of the square of the number of anyonic charges). Once again, the optimal approach is a series of pairwise contractions in which we contract together two anyonic tensors at a time, considering only the portion of labelled fusion diagram associated with this pair and the degeneracy tensors associated with this pair. To see how this saving arises, consider the simple contraction shown in \fref{fig:paircont2}, and let us na\"\i{}evely evaluate tensor $(AB)$ without first eliminating the loop.
Assuming for the moment that no shortcut exists for determining valid charge pairs $b,c$ consistent with given values of $a$ and $d$, one must iterate over all possible charges for each of $b$ and $c$, checking consistency at the vertices $a,b,c$ and $b,c,d$. Matters are simplified slightly by applying global conservation of charge to show that $a=d$, but still, if the anyon model admits $n_q$ charges, for each value of $a$ one must iterate over $(n_q)^2$ labellings of the fusion diagram. If $A$ and $B$ are part of a more extensive network, then in general the number of labellings of this network which must be checked will scale as $\mrm{O}[(n_q)^a],~a>2$. Contracting $A$ with $B$ to yield a single tensor $(AB)$ bearing only charge indices $a$ and $d$ thus reduces the number of labellings for the entire network by a factor of $(n_q)^2$ to $\mrm{O}[(n_q)^{a-2}]$. This saving significantly reduces computational cost, and if using a precomputation scheme to avoid repeatedly calculating quantities such as the unitary matrices arising from $F$-moves and braids,\cite{singh2011} then this approach also significantly reduces the amount of storage space which precomputed data requires.

This example is a little unrealistic, because the same saving can be achieved by identifying iteration over values of $b$, $\mu_b$, and $\mu_c$ with iteration over $\mu_a$, equivalent to eliminating the loop diagrammatically. This ceases to be universally possible, however, for loops involving the fusion trees of three or more tensors. Consequently, where a fusion diagram includes two or more loops involving three of more tensors apiece, the use of pairwise contractions ensures that these loops are eliminated one at a time for a saving in the scaling of both computational cost as a function of $n_q$, and precomputation storage, again as a function of $n_q$.

For some anyon models, there exist shortcuts which may cause the cost savings to be somewhat less pronounced than would be expected from the above example. Consider again the na\"\i{}eve calculation given above, this time specifically for an Abelian anyon model in the $\mbb{Z}_q$ series. In such a model one may immediately determine valid values of $c$ given $a$ and $b$, thus reducing the saving to a factor of $\mrm{O}(n_q)$ rather than $\mrm{O}[(n_q)^2]$. Similar cost savings may also be achieved in loops involving three or more tensors, but the scaling penalty associated with addressing the tensor network as a whole may never be entirely eliminated if the network includes two or more loops involving three or more tensors. For this reason, a pairwise approach to the contraction of tensors is deemed preferable.

One important question remains: Given the preferential nature of pairwise tensor contraction, what is the optimal sequence of pairwise contractions by which to contract a given anyonic tensor network? For normal tensors, this problem is known to be NP-hard.\cite{lam1997} Normal tensors may be considered to be a special case of anyonic tensors, with %
only the vacuum charge, trivial fusion rules, and trivial $F$-moves and braiding, and so it follows that this problem is NP-hard for networks of anyonic tensors as well. The best known approach for finding optimal contraction sequences is a brute-force computer search, with the state-of-the-art search for normal tensors being the \texttt{netcon} algorithm given in \rcite{pfeifer2014a}. Manual searches relying on human intuition may also be effective.
For practical purposes we will assume that if \texttt{netcon} identifies a contraction sequence as being optimal for normal tensors, then it will also be optimal or near-optimal for anyonic tensors. Certainly, it is simple to confirm that for DMRG these sequences display the optimal scaling of $\mrm{O}(D^3)$ in terms of the DMRG refinement parameter $D$ (the bond dimension of the MPS), and so they are sufficiently close to optimal for practical purposes.

\subsubsection{Anyonic tensor trace\label{sec:ttrace}}

\begin{figure}
\includegraphics[width=\columnwidth]{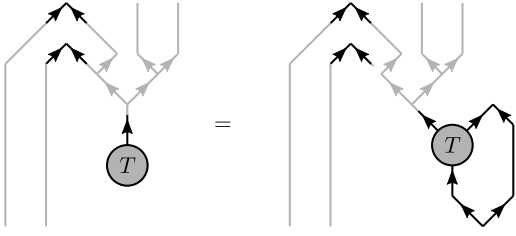} %
\caption{Graphical representation of tracing two indices on a three-index tensor, $T^\beta = T^{\beta\alpha}_\alpha$.\label{fig:ttrace}}
\end{figure}

For completeness we now describe one further single-tensor operation, namely the anyonic tensor trace. To evaluate a trace, it is necessary that the traced leaves be collected into one upper and one lower greek index, and any leaves not participating in the trace be collected (using bends if necessary) into a third greek index, which may be either upper or lower (see \fref{fig:ttrace}).
Note that upper leaves are paired with lower leaves, so all leaves involved in the trace are connected by a line making a pair of consecutive bends. We note that:
\begin{enumerate}
\item After tracing, the untraced third greek index is a single index and thus the resulting object has non-trivial entries only where the charge label in this greek index is the identity.
\item The fusion diagram associated with tracing is therefore always a loop, even when a third index is present. It is normalised according to the diagrammatic isotopy convention, and hence is associated with a factor of $d_a$.
\end{enumerate}
It therefore follows that the trace in \fref{fig:ttrace} may be evaluated according to
\begin{equation}
T^\beta = T^{\beta\alpha}_\alpha~\raisebox{-8.5pt}{\includegraphics[width=0.3in]{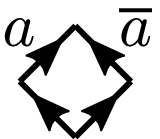}}
\, = \sum_{a,\mu_a} \,T^{(b,\mu_b)(a,\mu_a)}_{(a,\mu_a)}\, d_a.\label{eq:ttrace}
\end{equation}

\section{Finite DMRG on the disc\label{sec:finiteADMRG}}

\subsection{Matrix Product State Ansatz\label{sec:aMPS}}

\subsubsection{Matrices vs. tensors}

Two common notations for the Matrix Product State (MPS) Ansatz exist. We review these notations briefly in the context of a conventional MPS before proceeding to the anyonic case.

Let $\mc{L}$ be a lattice of $n$ sites with local dimension $d$ and local basis $\{|j\ra\},~1\leq j\leq d$. In the first notation, a family of $d$ matrices is associated with each lattice site. At lattice site $i$ we will denote these $d$ matrices $A^{[i]}_j,~1\leq j\leq d$. The probability amplitude associated with a state $|j_1j_2\ldots j_n\ra$ on lattice $\mc{L}$ is then given by the matrix product
\begin{equation}
c_{j_1j_2\ldots j_n} = A^{[1]}_{j_1}A^{[2]}_{j_2}\ldots A^{[n-1]}_{j_{n-1}}A^{[n]}_{j_n}\label{eq:MPSA}
\end{equation}
where the matrix at each site is selected according to the basis element $|j_1j_2\ldots j_n\ra$.
Strictly speaking, for open boundary conditions all matrices $A^{[1]}_j$ are row vectors and all matrices $A^{[n]}_j$ are column vectors so that $c_{j_1j_2\ldots j_n}$ is a number.

In the second notation it is recognised that a family of matrices, enumerated by an index $j$, is in fact equivalent to a three-index tensor $\Gamma^a_{bj}$ where $j$ is the index selecting the matrix in \Eref{eq:MPSA} and $a$ and $b$ are the row and column indices of this matrix respectively. In this notation, \Eref{eq:MPSA} becomes
\begin{equation}
c_{j_1j_2\ldots j_n} = \left(\Gamma^{[1]}\right)_{aj_1} \!\left(\Gamma^{[2]}\right)^a_{bj_2} \!\ldots \left(\Gamma^{[n-1]}\right)^c_{dj_{n-1}} \!\left(\Gamma^{[n]}\right)^d_{j_n},
\end{equation}
where repeated indices are summed. Again note that $\Gamma^{[1]}$ and $\Gamma^{[n]}$ have one fewer index, indicating that for a given value of $j_1$, $(\Gamma^{[1]})_{aj_1}$ is a row vector whose entries are enumerated by $a$ and for a given value of $j_n$, $(\Gamma^{[n]})^d_{j_n}$ is a column vector whose entries are enumerated by $d$.

As we have developed a formalism for anyonic tensors, we find the latter formulation to be better suited to our current circumstances.

\subsubsection{Graphical representation of the anyonic MPS}

Having established a formalism for anyonic tensor networks, we may now write down the anyonic version of the Matrix Product State (MPS) Ansatz with open boundary conditions. This is shown for open boundary conditions in \fref{fig:aMPS}, and differs from the conventional graphical representation of the MPS as follows:
\begin{figure}
\includegraphics[width=\columnwidth]{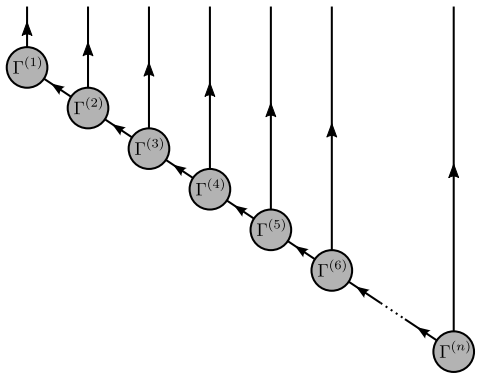} %
\caption{Anyonic Matrix Product State Ansatz for $n$ sites with open boundary conditions.\label{fig:aMPS}}
\end{figure}%
\begin{itemize}
\item Because anyonic states are represented by fusion trees with all leaves at the top, the open indices of the anyonic MPS are upward-pointing rather than downward-pointing.
\item There are no horizontal index lines: All lines on a fusion tree have an upper and a lower end.
\item All index lines carry orientation arrows (though these may be omitted for models where charges are self-dual). For our convenience, we choose these all to be upward-pointing.
\item On a conventional MPS, the ability of the Ansatz to faithfully represent an arbitrary state is constrained by the dimensions of the indices connecting the $\Gamma$ tensors, which are restricted to be no greater than some refinement parameter $D$. For an anyonic MPS one must also specify which charges may appear on a given index, and divide the total dimension $D$ between these charges.
\end{itemize}

To perform DMRG, we will also require a graphical representation of a Hamiltonian. In the first instance we will consider a Hamiltonian which may be written as a sum of local terms. For specificity we choose these terms to be two-site, having a graphical representation 
\begin{equation}
h_{ij} = \raisebox{-30pt}{\includegraphics[width=0.3in]{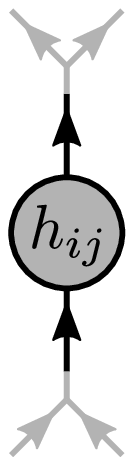}}\,. 
\end{equation}
We will subsequently extend our treatment to %
also include Matrix Product Operator (MPO) Hamiltonians in \sref{sec:MPO}.

\subsection{DMRG algorithm with open boundary conditions\label{sec:OBC}}

We consider now how the DMRG algorithm may be applied to the anyonic MPS of \fref{fig:aMPS} to compute an approximation to the ground state of a Hamiltonian. As in conventional DMRG, the anyonic MPS is converged towards a representation of the ground state by means of a process of variational optimisation. In one step of the optimisation process any number of consecutive $\Gamma$ tensors may be updated simultaneously, though for reasons subsequently made clear we recommend at least two. The first optimisation is therefore of $\Gamma_1$ and $\Gamma_2$, followed by $\Gamma_2$ and $\Gamma_3$, then $\Gamma_3$ and $\Gamma_4$, and so on. Upon reaching the end of the chain, the process reverses, first optimising $\Gamma_{n-1}$ and $\Gamma_n$, then $\Gamma_{n-2}$ and $\Gamma_{n-1}$, and so forth.

\begin{figure*}
\includegraphics[width=492pt]{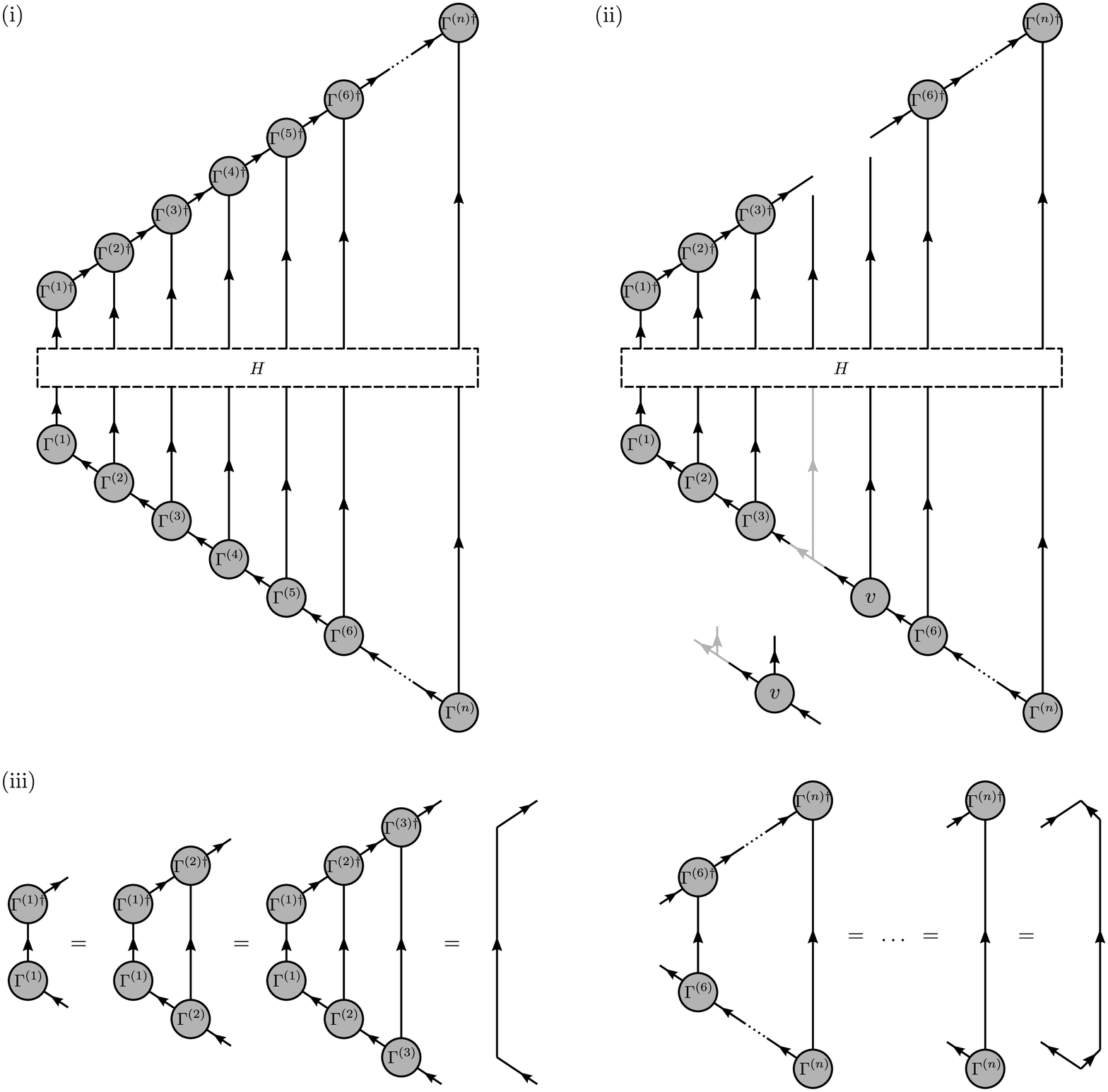} %
\caption{DMRG algorithm part~1: (i)~Graphical representation of $\la\psi|\hat{H}|\psi\ra$, with the Hamiltonian represented schematically by a dotted box. (ii)~Tensors $\Gamma_4$ and $\Gamma_5$ are replaced by a tensor $v$, whose structure is shown in the inset. (iii)~Careful preparation ensures that tensors $\Gamma_1$, $\Gamma_2$, and $\Gamma_3$ satisfy some useful identities, as do $\Gamma_6,\ldots,\Gamma_n$.\label{fig:DMRG1}}
\end{figure*}

To understand the optimisation procedure, let us consider a specific pair, $\Gamma_4$ and $\Gamma_5$, while iterating from left to right. We are seeking a choice of $\Gamma_4$ and $\Gamma_5$ which minimises the value of $\la\psi|\hat H|\psi\ra$, represented schematically by \fref{fig:DMRG1}(i).
\begin{figure*}
\includegraphics[width=492pt]{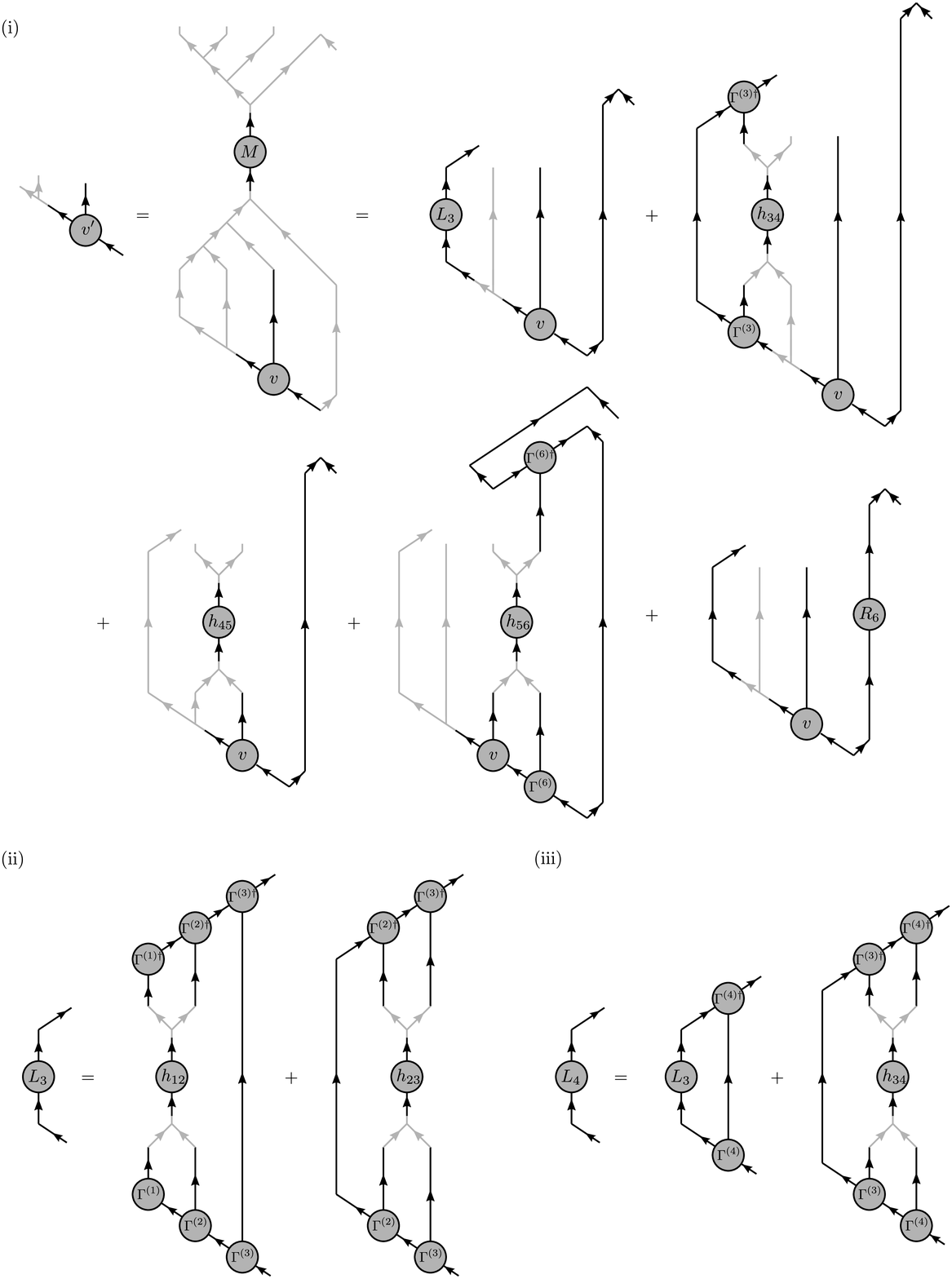} %
\caption{DMRG algorithm part~2: (i)~Construction of the product $Mv$. The terms have been simplified using the identities of \pfref{fig:DMRG1}(iii). %
(ii)~Construction of operator $L_3$ in diagram~(i). Construction of $R_6$ proceeds analogously on the right. (iii)~After updating $\Gamma_4$ and $\Gamma_5$, tensor $L_4$ is constructed in preparation for the update of $\Gamma_5$ and $\Gamma_6$.\label{fig:DMRG2}}
\end{figure*}%
Deleting $\Gamma_4$ and $\Gamma_5$ from this figure, we first seek a tensor $v$ with four leaves which may be inserted in their place [\fref{fig:DMRG1}(ii)], and which maximises the value of $\la\psi|\hat H|\psi\ra$ subject to the constraint $\la\psi|\psi\ra=1$. If we reshape $v$ and $v^\dagger$ as vectors $v^\alpha$ and $v^\dagger_\alpha$ and the rest of the tensor network as a matrix $M^\alpha_\beta$ [similar in principle to the way that $o$ and $S$ were reshaped as vector and matrix in \fref{fig:superop2}(i)], and we define $N^\alpha_\beta$ as the matrix obtained when $\hat H$ is replaced by the identity operator, then this takes the form of a generalised eigenvalue problem
\begin{equation}
Mv = \lambda Nv
\end{equation}
where we are looking for the choice of $v$ which minimises the value of $\lambda$ for the specified $M$ and $N$. A little preparation in getting to this point ensures that the $\Gamma$ tensors satisfy the identities given in \fref{fig:DMRG1}(iii)---we shall see how this is enforced when we obtain the updated versions of $\Gamma_4$ and $\Gamma_5$---and the problem then simplifies to an eigenvalue problem
\begin{equation}
Mv=\lambda v\label{eq:EVprob}
\end{equation}
where we seek the vector $v$ which minimises $\lambda$. The construction of the product $Mv$ in a manner which preserves the structure of $v$ is given in \fref{fig:DMRG2}(i), where $L_3$ is constructed as per \fref{fig:DMRG2}(ii) and $R_6$ is constructed analogously using tensors from the region of the Ansatz to the right of $v$.

\begin{figure}
\includegraphics[width=\columnwidth]{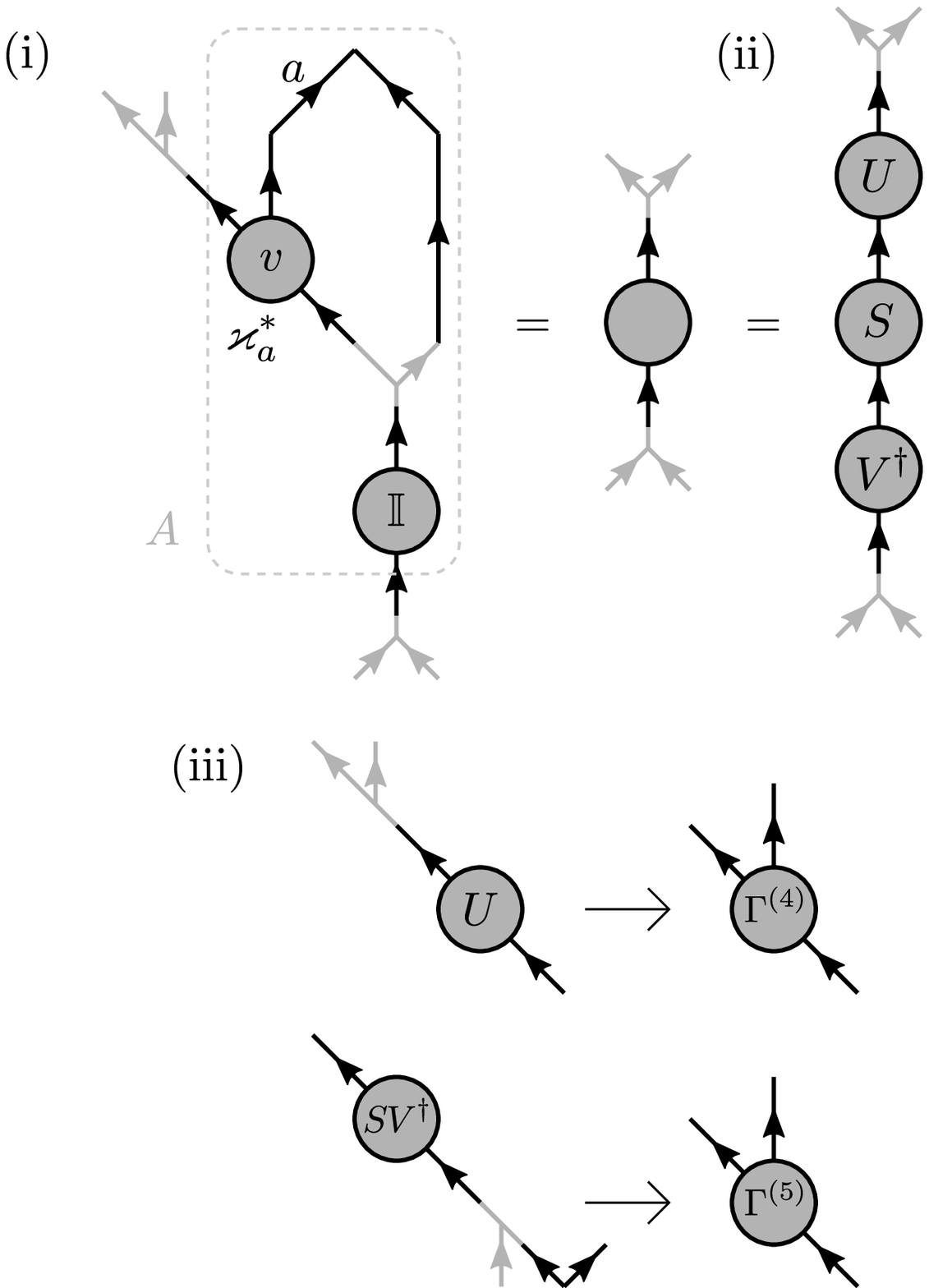}
\caption{(i)~Reshaping of $v$ into a matrix. (ii)~Singular value decomposition of $v$ into $U$, $S$, and $V^\dagger$. (iii)~Construction of updated $\Gamma^{(3)}$ and $\Gamma^{(4)}$ from $U$ and $SV'$.\label{fig:svd}}
\end{figure}

Obtaining an eigenvector $v$ using the Lanczos method, we reshape it into a matrix as shown in \fref{fig:svd}(i) and then perform a singular value decomposition [\fref{fig:svd}(ii)]. 
Matrix $S$ is diagonal, containing the positive real singular values, and satisfies the normalisation condition
\begin{equation}
S^\alpha_\beta S^\beta_\alpha=1,\label{eq:norm}
\end{equation}
or equivalently, if the diagonal entries of $S$ in charge sector $q$ are denoted $s^1_q,\ldots,s^n_q$,
\begin{equation}
\sum_q \sum_i (s^i_q)^2 d_q = 1\label{eq:normsum}
\end{equation}
where the factor of $d_q$ arises from the loop in the diagram, as per Secs.~\ref{eq:summationcaveat} and \ref{sec:ttrace}.

In general, the number of entries in $S$ may now exceed the refinement parameter $D$. Tensor $S$ is therefore truncated,
retaining the $D$ non-zero entries %
which contribute the largest quantities to the sum in \Eref{eq:normsum}, and the indices on $U$ and $V^\dagger$ which connect to $S$ are truncated likewise. The truncated matrix $S_\mrm{trunc}$ is then normalised according to
\begin{equation}
\left(S_\mrm{trunc}\right)^\alpha_\beta \rightarrow \frac{\left(S_\mrm{trunc}\right)^\alpha_\beta}{\left(S_\mrm{trunc}\right)^\gamma_\delta\left(S_\mrm{trunc}\right)^\delta_\gamma}\label{eq:renorm}
\end{equation}
such that \Eref{eq:norm} is again satisfied. If $D_q$ entries are retained in charge sector $q$, we say that charge sector $q$ is of dimension $D_q$ on the indices connecting $U$ with $S$ and $S$ with $V^\dagger$. %

As we are iterating from left to right, the diagonal matrix $S_\mrm{trunc}$ is now absorbed into $V^\dagger$ and the new $\Gamma_4$ and $\Gamma_5$ are constructed from $U$ and $SV^\dagger$ as shown in \fref{fig:svd}(iii). Construction of $\Gamma_4$ entirely from the unitary matrix $U$ ensures that when updating $\Gamma_5$ and $\Gamma_6$, the identity given in \fref{fig:DMRG1}(iii) now extends to $\Gamma_4$ as well.

We also note that the index which connected tensor $U$ with tensor $S$ now connects the new $\Gamma_4$ with the new $\Gamma_5$. The dimensions $D_q$ of the different charge sectors on this index are determined entirely by the relative contributions of the singular values to \Eref{eq:normsum}, and these dimensions may be different to those which were exhibited by the old $\Gamma_4$ and $\Gamma_5$. This opportunity to optimise the dimensions of the different charge sectors as a part of the variational update procedure is the reason why we recommend updating at least two $\Gamma$-tensors at once.

In preparation for the next update step, tensor $L_4$ is now computed from $L_3$ and $\Gamma_4$ as shown in \fref{fig:DMRG2}(iii), though tensor $L_3$ is also retained in memory as it will be needed again when the next pass, which iterates from right to left, updates tensors $\Gamma_4$ and $\Gamma_5$. Similarly, while iterating from left to right we employed tensor $R_6$ in \fref{fig:DMRG2}(i), which was constructed after updating tensors $\Gamma_5$ and $\Gamma_6$ on the last pass which was performed from right to left.

\subsection{Matrix Product Operator Hamiltonians\label{sec:MPO}}

\begin{figure*}
\includegraphics[width=492pt]{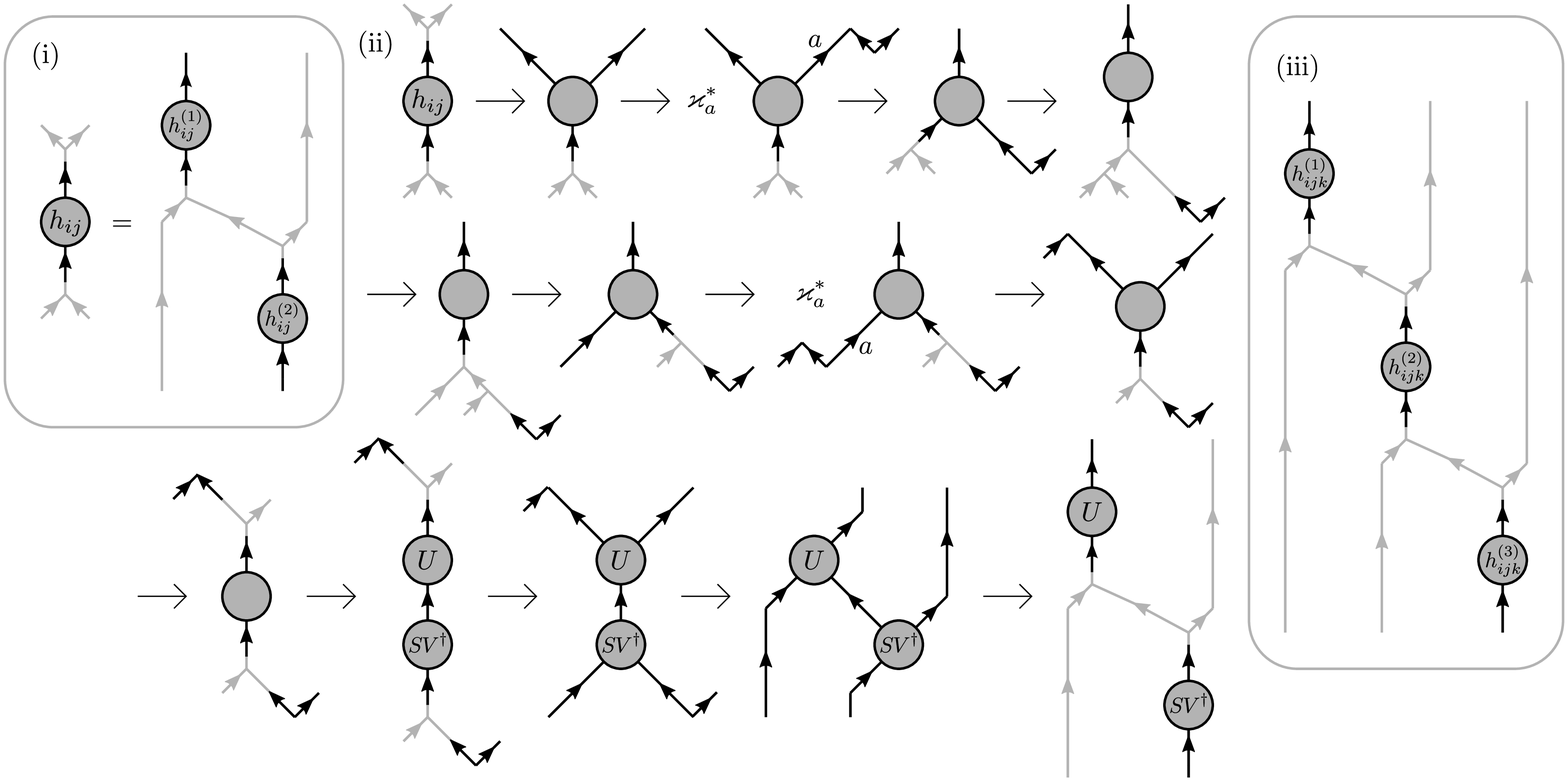}
\caption{(i)~Matrix Product Operator (MPO) decomposition of a two-site anyonic operator. (ii)~Calculation of an example MPO decomposition of a two-site anyonic operator. (iii)~MPO decomposition of a three-site anyonic operator.\label{fig:MPO}}
\end{figure*}

We now introduce Matrix Product Operator (MPO) decompositions for individual terms in the Hamiltonian.\cite{mcculloch2007} Considering first a two-site operator $\hat h_{ij}$ such as was used in \sref{sec:OBC}, by collecting together the left two indices and the right two indices and performing a singular value decomposition this operator may be rewritten in the form of \fref{fig:MPO}(i), as shown in \fref{fig:MPO}(ii).
This decomposition may be extended to larger operators through the use of repeated singular value decompositions, each separating off one site at a time to yield structures such as the three-site MPO operator shown in \fref{fig:MPO}(iii). Extrapolation to four or more sites is straightforward.

\begin{figure*}
\includegraphics[width=492pt]{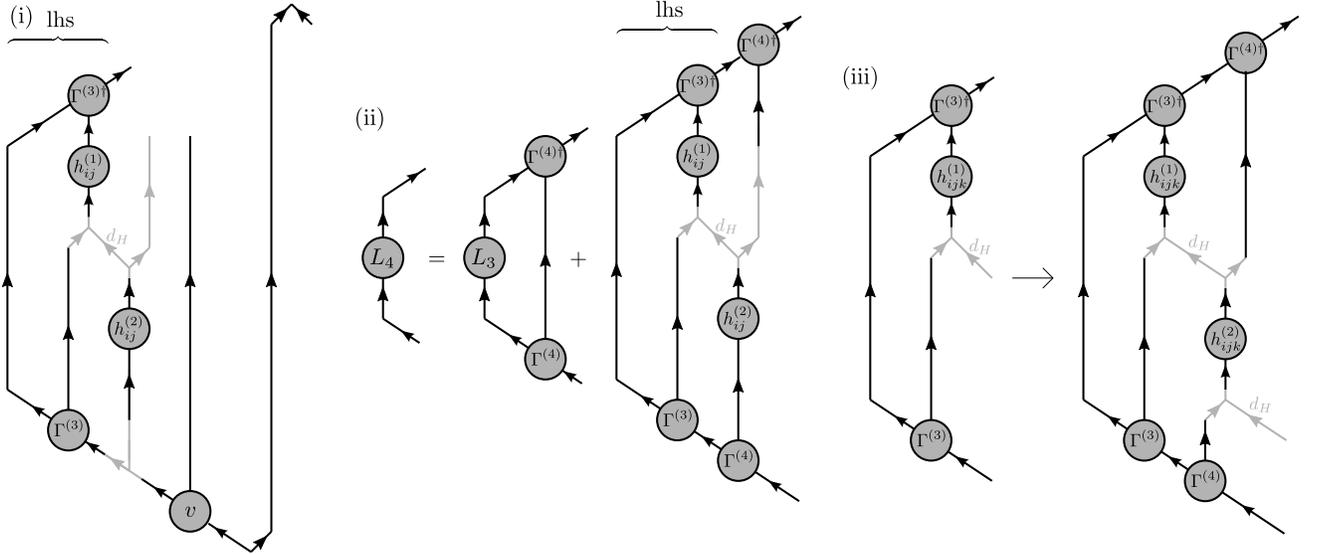}
\caption{(i)~Diagram equivalent to the $h_{34}$ term in \fref{fig:DMRG2}(i), with the Hamiltonian term expressed in MPO form. The internal bond of the MPO is indicated as being of dimension~$d_H$, while physical sites are of dimension~$d$ and MPS bond indices are of dimension~$D$. (ii)~Calculation of $L_4$ with MPOs. Note the re-use of the region marked ``lhs'' in diagram~(i). (iii)~If working with a Hamiltonian having terms involving more than two sites, the ``lhs'' for the term acting on sites~3 and up during optimisation of $\Gamma_4$ and $\Gamma_5$ may be extended to obtain the ``lhs'' for the term acting on sites~3 and up during optimisation of $\Gamma_5$ and $\Gamma_6$ by incorporating the next portion of the MPO operator as shown.\label{fig:MPOexpec}}
\end{figure*}

The advantages of working with MPO operators are readily apparent, even for two-site operators. For example, when applying matrix $M$ to vector $v$ in \fref{fig:DMRG2}(i), the term involving $\hat h_{34}$ now appears as shown in \fref{fig:MPOexpec}(i). 
If the dimension of the index labelled $d_H$ is less than that of $d\times d$, where $d$ represents a physical lattice site, then the cost of evaluating this diagram is reduced compared to \fref{fig:DMRG2}(i). In addition, when computing $L_4$ the left-hand side of \fref{fig:MPOexpec}(i) (labelled lhs) may be re-used, as shown in \fref{fig:MPOexpec}(ii). This latter advantage becomes more pronounced when working with Hamiltonians whose terms span more than two sites, with only a portion of each operator overlapping the sites for which the $\Gamma$-tensors are being updated. For example, suppose that the Hamiltonian is made up of a sum of three-site MPO terms. One may take the region labelled lhs, which describes the leftmost site of a term acting on sites~3 and up, and extend it as shown in \fref{fig:MPOexpec}(iii) to represent the \emph{two} leftmost sites of the same term acting on sites~3 and up during the update of tensors $\Gamma_5$ and $\Gamma_6$. Not only can we recycle our contraction of the region labelled lhs in \fref{fig:MPOexpec}(i) from one update round to the next, but within a given update round we have partially precomputed the contribution of one term in the Hamiltonian to the evaluation of $Mv$, and this precomputed portion remains the same from one iteration to the next.

Of course, sooner or later the final portion of an MPO is appended onto an lhs portion, and the object is then added into $L$, which contains all terms in the Hamiltonian lying entirely to the left of the $\Gamma$-tensors being updated. We may, however, cache the lhs objects for re-use when sweeping back from right to left---and needless to say, a similar treatment pertains to the portions of Hamiltonian terms which  extend to the right of the $\Gamma$-tensors being updated.

Depending on the construction of the Hamiltonian, it may be preferable to perform an MPO decomposition term by term, on collections of terms together, or possibly even on the Hamiltonian as a whole. However, provided $d_H\ll D$ the cost of the DMRG algorithm in all cases continues to scale as $\mrm{O}(D^3)$. More involved discussions of the role of MPO decompositions in DMRG may be found in Refs.~\onlinecite{schollwock2011} and \onlinecite{mcculloch2007}.

\subsection{Periodic boundary conditions\label{sec:PBC}}

A form of the DMRG algorithm for non-anyonic systems with periodic boundary conditions and a computational cost scaling as $\mrm{O}(D^3)$ is given in \rcite{pippan2010}. Although less accurate for a given value of $D$ than finite or infinite DMRG with open boundary conditions, this algorithm may be useful when seeking to avoid edge effects and we therefore describe its adaptation to anyonic systems here.

\begin{figure}
\includegraphics[width=\columnwidth]{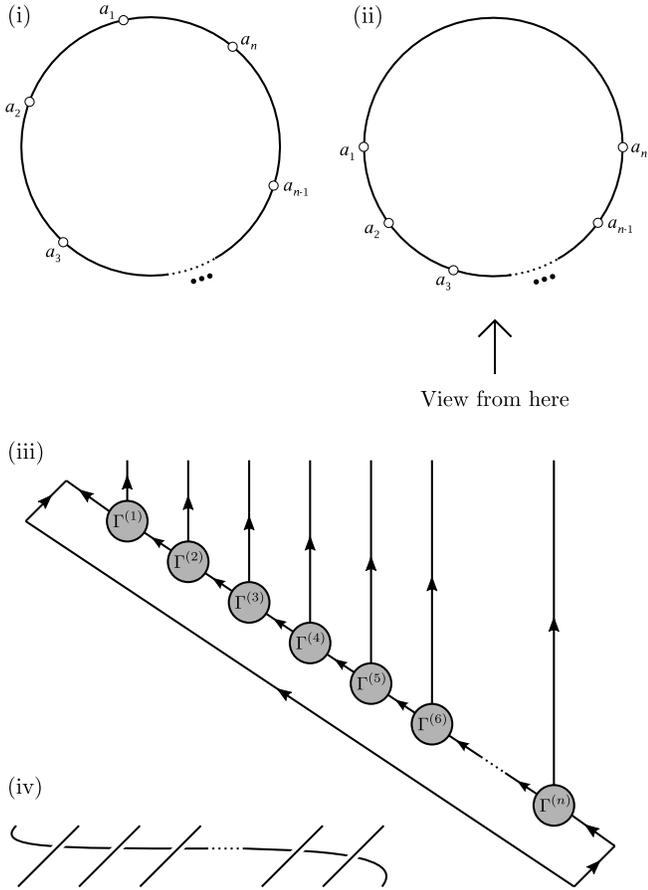}
\caption{(i)~Anyons on a circular lattice on the surface of the disc. (ii)~Adopting the viewpoint shown, deform the disc so that all anyons are brought to the front of the ring. (iii)~This system may be described using a periodic MPS Ansatz, and the geometry specified in diagram~(ii) reveals that the periodic translation operator acts as shown in diagram~(iv). (See also \protect{\rcite{pfeifer2012a}}.)%
\label{fig:anyonsonring}}
\end{figure}

We begin by specifying the construction of the anyonic MPS for a ring of anyons on the disc.
First, let the anyons lie on a ring as shown in \fref{fig:anyonsonring}(i), and then, for our convenience, let us deform the disc so that all of these anyons are brought to the front side of the ring [\fref{fig:anyonsonring}(ii)].
An MPS Ansatz for this system of anyons is given in \fref{fig:anyonsonring}(iii), and it follows from the construction specified in \fref{fig:anyonsonring}(i)-(ii) that the periodic translation operator for this Ansatz appears as shown in \fref{fig:anyonsonring}(iv). The action of the periodic translation operator on the MPS Ansatz may be evaluated using the same method as was employed in \rcite{pfeifer2012a} for the periodic translation operator on the torus. Using this method, we see that the action of \fref{fig:anyonsonring}(iv) on the periodic anyonic MPS is to cyclically permute the $\Gamma$-tensors of the MPS, and also to multiply each element of the outside tensor [$\Gamma_n$ in \fref{fig:anyonsonring}(iii)] by a phase factor of $R^{a\overline{a}}_\mbb{I}$ where $a$ is the corresponding charge on the physical index of the tensor for that element.

\begin{figure}
\includegraphics[width=\columnwidth]{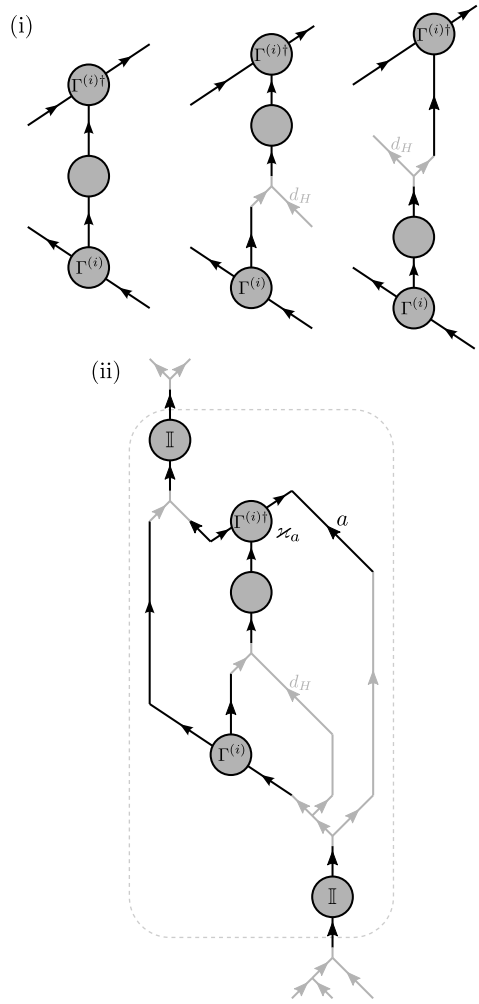}
\caption{(i)~Transfer ``matrices'' in the periodic DMRG algorithm. (Strictly, transfer networks.) (ii)~Example showing how these anyonic tensor networks can be expressed in matrix form through the application of identity operators.\label{fig:PBC_TMs}}
\end{figure}

Now that we have an anyonic version of the periodic MPS Ansatz, extension of the periodic DMRG algorithm to anyonic systems is relatively straightforward. At the heart of this algorithm is a transfer matrix approach, where each term in the Hamiltonian is written in the form of an MPO and for each term, each site of the lattice is then associated with a transfer ``matrix'' having one of the forms given in \fref{fig:PBC_TMs}(i).
[Strictly speaking, the objects shown in \fref{fig:PBC_TMs}(i) are transfer tensor networks, but they may be expressed in matrix form as shown in \fref{fig:PBC_TMs}(ii).]
The algorithm exploits the fact that when a long chain of identical transfer matrices is multiplied together, only a relatively small number of singular values make a significant contribution to the end result. Consequently, one can approximate the transfer matrix (which has dimensions ranging between $D^2\times D^2$ and $D^2d_H\times D^2d_H$ depending on the MPO contribution) by a product denoted $UdV$ where $UdV$ has the same dimensions as the original transfer matrix, say $D^2\times D^2$ for example, but $d$ is much smaller, perhaps $4\times 4$, and $U$ and $V$ are then $D^2\times 4$ and $4\times D^2$ respectively.\footnote{Where dimensions of indices are given as products such as \protect{$D^2$} and \protect{$D^2d_H$}, these are to be understood as representing anyonic fusions rather than conventional multiplication; thus, for example, \protect{$D^2$} denotes the dimension of the index obtained on fusing together two MPS bond indices of dimension $D$, taking into account the charges represented, the tensor network structure, and the fusion rules.}

Crucial to this calculation is the ability to compute the approximate singular value decomposition $UdV$ for a cost of $\mrm{O}(D^3)$. An algorithm for doing so is provided in \rcite{pippan2010}, and begins by multiplying the transfer matrix (which \citeauthor{pippan2010} denote $M$, but we shall write as $T$) by a $D^2\times p$ (or $D^2d_H\times p$) matrix $x$ where $p$ is the number of singular values to be retained. This algorithm functions unchanged for transfer networks put in matrix form as per \fref{fig:PBC_TMs}, though we note that it is necessary to decide how the singular values will be divided among the different charge sectors when constructing $x$.

Finally, we note a couple of practical considerations: In the preferred form of the periodic DMRG algorithm described at the end of \rcite{pippan2010}, the periodic MPS is divided into thirds and iteration over the $\Gamma$-tensors is performed one third at a time. We observe that it is probably easiest to use the periodic translation operator to rotate the third of the MPS being updated into the centre of the Ansatz, though this is not essential. Finally, we observe that the expression for $Mv$ [where $M$ is the matrix of the eigenvalue problem, as in \Eref{eq:EVprob}, and not the transfer matrix, as in \rcite{pippan2010}] involves tensor traces. These are evaluated as described in \sref{sec:ttrace}.

\section{Finite DMRG on the torus\label{sec:torus}}

\begin{figure}
\includegraphics[width=\columnwidth]{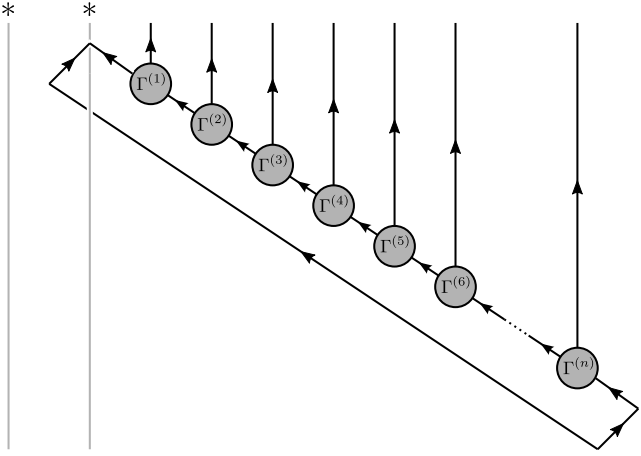}
\caption{Matrix Product State Ansatz for a ring of anyons encircling one cycle of the torus (i.e.~without twist). For an explanation of the $*$ notation, see \protect{\rcite{pfeifer2012a}}.\label{fig:anyonsontorus}}
\end{figure}

Generalisation from periodic boundary conditions on the disc to periodic boundary conditions on the torus %
may be achieved by following the prescriptions for surfaces of higher genus given in \rcite{pfeifer2012a}. We consider here the simplest such scenario, where a ring of lattice sites encircles the torus around one cycle only. More complex two-dimensional scenarios will be considered in forthcoming papers. In this scenario, the Ansatz of \fref{fig:anyonsonring}(iii) for periodic boundary conditions on the disc is replaced by \fref{fig:anyonsontorus} for anyons on the torus. 
Once again, one applies the periodic DMRG algorithm to a third of the $\Gamma$-tensors at a time, and it may be convenient to use the periodic translation operator to place those tensors at the centre of the Ansatz. The periodic translation operator now takes the form
\begin{equation}
\raisebox{-12pt}{\includegraphics{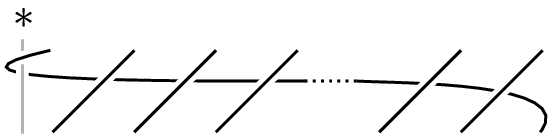}}
\end{equation}
and its action is again to cyclically permute the lattice sites and multiply by a factor $R^{a\overline{a}}_\mbb{I}$. The key difference lies in the definition of the inner product between a bra and a ket. For anyons on the disc this is given by \Eref{eq:innerprod}, with the charge on index $\alpha$ necessarily being $\mbb{I}$. For a ring of anyons on the torus, the corresponding inner product is
\begin{equation}
\raisebox{-30pt}{\includegraphics[width=1.25in]{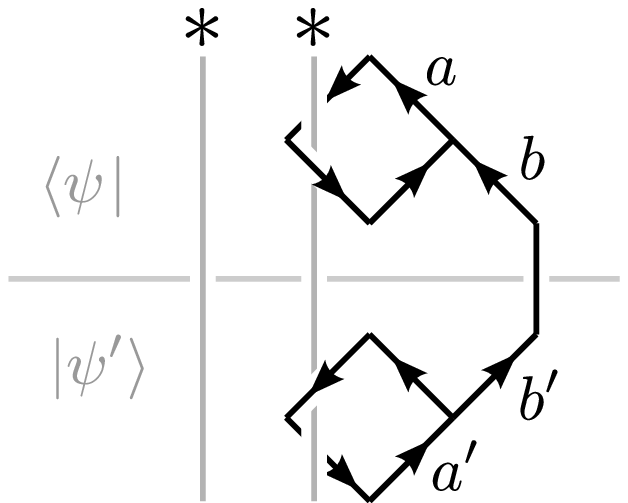}}=\delta_{aa'}\delta_{bb'}\,\sqrt{d_{b}}\,d_{a}^2\label{eq:torusinnerprod}
\end{equation}
with the tensor generalisation giving rise to the normalisation condition for a state:
\begin{equation}
\begin{split}
\raisebox{-30pt}{\includegraphics[width=1.25in]{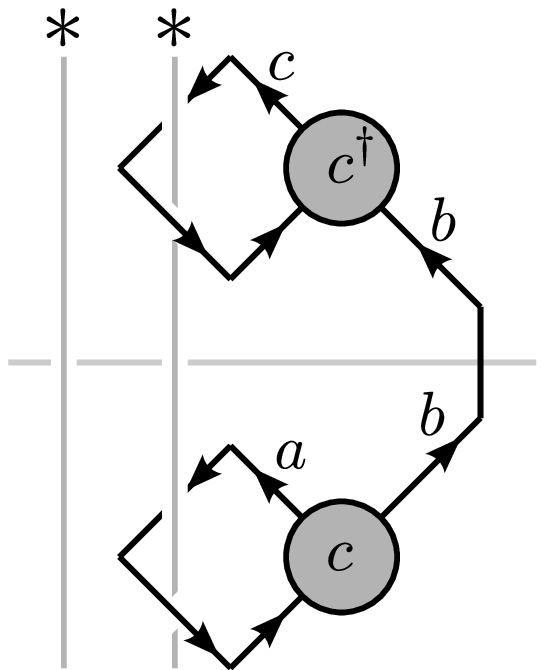}}&\begin{split}
=\sum_{\substack{a,\mu_a\\b,\mu_b\\c,\mu_c}} &c^{(a,\mu_a)(b,\mu_b)}_{(a,\mu_a)} c^{\dagger(c,\mu_c)}_{\phantom{\dagger}(c,\mu_c)(b,\mu_b)}\\
&\times \delta_{ac}\,\delta_{\mu_a\mu_c}\sqrt{d_b}\,d_a^2
\end{split}
\\~\\
&=1,
\end{split}\label{eq:torusstatenorm}
\end{equation}
where any charges are permitted on the indices provided the corresponding vertices within the tensors are consistent with the anyonic fusion rules. [Note that Eqs.~\eref{eq:torusinnerprod}--\eref{eq:torusstatenorm} have been expressed in a diagrammatic form consistent with \rcite{pfeifer2012a}, and that to apply this expression to a state constructed from $\Gamma$-tensors requires vertical bending. For reference, if $c$ were constructed by contracting a loop of $\Gamma$-tensors then index $\alpha=(a,\mu_a)$ would be of dimension $D$, the dimension of the MPS bond, and index $\beta=(b,\mu_b)$ would be of dimension $d^n$, the dimension of $n$ physical sites.]

\section{Example}

We include one simple example to illustrate the capabilities of anyonic DMRG. The Golden Chain\cite{feiguin2007} is one of the best understood non-Abelian anyonic models, comprising a chain of fixed Fibonacci anyons ($\tau$) subject to a nearest-neighbour interaction which favours pairwise fusion into either the $\tau$ channel (ferromagnetic interaction) or the vacuum channel (antiferromagnetic interaction). This model has been studied extensively using exact diagonalisation,\cite{feiguin2007,pfeifer2012,pfeifer2012a} Time-Evolving Block Decimation,\cite{singh2014} valence bond Monte Carlo,\cite{tran2010} and the Multi-Scale Entanglement Renormalisation Ansatz.\cite{pfeifer2010} DMRG has previously been used to compute the central charge,\cite{feiguin2007,trebst2008a} and we apply it now to calculation of the ground state energy of finite chains.

\begin{figure}
\includegraphics[width=\columnwidth]{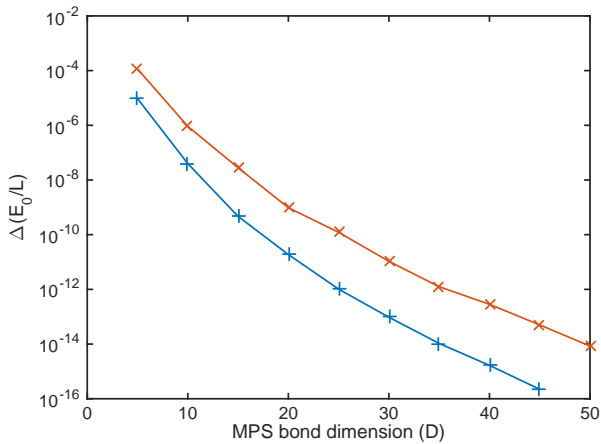}
\caption{COLOUR ONLINE: We define the quantity \protect{$\Delta (E_0/L)$} as the difference between the ground state energy per site for a given MPS bond dimension \protect{$D$}, and the value in the large-\protect{$D$} limit%
. In this graph we plot \protect{$\Delta (E_0/L)$} as a function of $D$ for $L=50$ ferromagnetic (\protect{$\times$}) and antiferromagnetic (\protect{$+$}) Fibonacci anyon chains. We encounter the limit of numerical precision at $D=50$ for the antiferromagnetic chain and at $D=55$ for the ferromagnetic chain.
\label{fig:results1}}
\end{figure}

\begin{figure}
\includegraphics[width=\columnwidth]{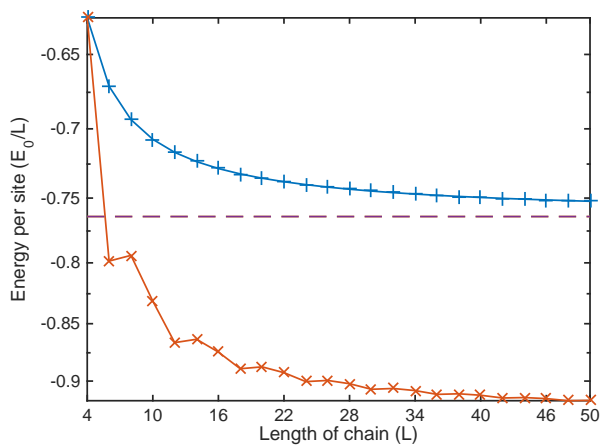}
\caption{COLOUR ONLINE: Energy per site for ferromagnetic (\protect{$\times$}) and antiferromagnetic (\protect{$+$}) Fibonacci anyon chains as a function of chain length $L$, with fixed MPS bond dimension $D=233$%
. The $L\rightarrow\infty$ limit for the antiferromagnetic model is shown as a dashed horizontal line.\label{fig:results2}}
\end{figure}

Both the ferromagnetic and antiferromagnetic Hamiltonians are critical, and thus interest in this model has concentrated primarily on the infinite chain. For the present paper, however, we are restricted to finite systems. %
In Figs.~\ref{fig:results1}--\ref{fig:results2} we therefore study the behaviour of the ground state energy per site $(E_0/L)$ %
as a function of refinement parameter $D$ and chain length $L$.
In \fref{fig:results1} we 
see that the ground state energy rapidly converges as a function of $D$. The results obtained are stable to over fourteen decimal places by $D=55$, essentially corresponding to the accuracy limit of our numerical eigensolver. Meanwhile for fixed $D=233$ we see that the energy per site continues to show a strong dependency on chain length, indicating significant finite size effects, especially for the ferromagnetic coupling. As we are essentially independent of $D$ in this regime, and
the energy per site of the antiferromagnetic chain smoothly approaches the $L\rightarrow\infty$ limit for even $L$, we perform a polynomial fit of $E_0/L$ as a function of $L^{-1}$ to extrapolate the energy per site of the antiferromagnetic model in the $L\rightarrow\infty$ limit.\\~\\
\begin{minipage}{\columnwidth}
\begin{ruledtabular}
\begin{tabular}{c|c|c}
Model & $E_0/L$ at $L=50$ & $E_0/L$ for $L\rightarrow\infty$\\
\hline
AFM Golden Chain & -0.752390661 & -0.763932014\\
\end{tabular}
\end{ruledtabular}
\end{minipage}\\~\\
If we compare the antiferromagnetic result with the exact value obtained in \rcite{tran2010}, 
\begin{equation}
\begin{split}
d&=2\cos\frac{\pi}{5}\\
\left.\frac{E_0}{L}\right|_{L\rightarrow\infty}&=\frac{d^2-4}{4d}\int_{-\infty}^\infty \rmd x\frac{\mrm{sech}(\pi x)}{\cosh\left(2x\arccos\frac{d}{2}\right)-\frac{d}{2}}\\
&\approx-0.763932023, 
\end{split}
\end{equation}
we find agreement to seven significant figures. For comparison, the TEBD algorithm on the infinite chain with $D=200$ is accurate to eight significant figures.\cite{singh2014} We therefore see that when energy per site is a smooth function of the length of chain, applying the DMRG algorithm to a series of finite chains and extrapolating to $L\rightarrow\infty$ is capable of yielding ground state energies with accuracy comparable to those obtained using infinite TEBD. As with conventional DMRG, one major advantage of the anyonic DMRG algorithm is the rapidity with which it converges: When generating \fref{fig:results2}, the ground state energies of the simulations were found in every case to have converged to a precision of between fourteen and fifteen significant figures after only a single iteration of the algorithm (one left-to-right sweep and one right-to-left sweep), being limited only by the accuracy of the numerical eigensolver employed.

We may also compare this result with the anyonic MERA of \rcite{pfeifer2010}. This paper analysed the Golden Chain using the scale-invariant 3:1~1D~MERA with $\chi=8$, divided between charge sectors as $\chi_\mbb{I}=3$, $\chi_\tau=5$. The computational cost of this simulation is of a similar order of magnitude to anyonic DMRG with $D=256$, as the cost of variationally optimising the 3:1~MERA without approximation scales as $\mrm{O}(\chi^8)$ while the cost of DMRG scales as $\mrm{O}(D^3)$. %
Using the 3:1~anyonic MERA with $\chi=[3,5]$ we find a ground state energy per site of
\begin{equation}
\left.\frac{E_0}{L}\right|^\mrm{MERA}_{L\rightarrow\infty}=-0.7639304\ldots,
\end{equation}
agreeing with the exact result to only five significant figures and requiring over $31,000$ iterations to reach this level of precision. %
Anyonic DMRG therefore provides a much faster route to the calculation of ground state energies than does the anyonic MERA; the primary advantage of the MERA calculation is that it gives direct access to the scaling operators and scaling dimensions of the critical theory. %

\section{Conclusion}

In this paper we have expanded upon the idea of anyonic tensors introduced in \rcite{pfeifer2010}, introducing a construction for these tensors which avoids evaluating numerical factors associated with whole tensor networks. We then used this construction to implement finite DMRG algorithms for open and periodic chains of anyons on the disc and torus.

The primary advantages of the DMRG algorithm are its speed and accuracy, coupled with the fact that it yields an explicit representation of the lowest-energy state identified.
We then applied the anyonic DMRG algorithm to
two of the most widely studied non-Abelian anyon models, the antiferromagnetic and ferromagnetic Golden Chains on the disc, and showed that the finite anyonic DMRG algorithm is capable of yielding highly accurate results for ground state energies on finite and infinite chains (the latter through finite size scaling). The accuracy of our results is comparable to that which can be obtained using anyonic TEBD and greater than can be obtained with an anyonic MERA, and they are obtained at substantially lesser computational cost. Anyonic DMRG consequently represents a highly effective technique for the numerical study of anyonic systems.

\begin{acknowledgments}
This research was supported in part by the ARC Centre
of Excellence in Engineered Quantum Systems (EQuS),
Project No. CE110001013.
\end{acknowledgments}

\appendix

\bibliography{AnyonicDMRG}

\end{document}